\documentclass[oupdraft]{bio}

\usepackage{url}
\usepackage{amsmath}
\usepackage{graphicx,psfrag,epsf}
\usepackage{enumerate}
\usepackage{booktabs}
\usepackage{color}
\usepackage{natbib}
\usepackage{paralist}
\usepackage{dsfont}

\newtheorem{property}[theorem]{Property}

\usepackage[framemethod=TikZ]{mdframed}
\RequirePackage[OT1]{fontenc}
\usepackage[utf8x]{inputenc}
\RequirePackage{amsthm}
\usepackage{relsize}
\usepackage{listings} 
\usepackage{epstopdf}
\usepackage[sc]{mathpazo}
\usepackage{ wrapfig}
\usepackage{dsfont}
\usepackage{graphicx}
\usepackage{caption}
\usepackage{adjustbox}
\DeclareCaptionFormat{myformat}{\hrulefill \\ #1#2#3}
\DeclareCaptionFont{small}{\footnotesize}
\DeclareCaptionFont{blue}{\color{blue}} 
\DeclareCaptionFont{bf}{\bfseries}
\usepackage{booktabs}

\usepackage{algorithm, eqparbox,array}
\usepackage{algpseudocode}
\usepackage{color}
\RequirePackage{xcolor}

\algnewcommand\INPUT{\item[\textbf{Input:}]}%
\algnewcommand\OUTPUT{\item[\textbf{Output:}]}%
\definecolor{mygray}{gray}{0.3}

\numberwithin{equation}{section}
\theoremstyle{plain}


\newcounter{theo}[section] 
\renewcommand{\thetheo}{\arabic{section}.\arabic{theo}}

\definecolor{light-gray}{gray}{0.95} 

\begin{document}

\title{Dimension constraints improve hypothesis testing for large-scale, graph-associated, brain-image data}

\author{TIEN VO, VAMSI ITHAPU, VIKAS SINGH,\\
 MICHAEL A. NEWTON$^\ast$ \\[4pt] 
\textit{Department of Biostatistics and Medical Informatics,
University of Wisconsin at Madison
610 Walnut Street,
Madison, WI
USA}
\\[2pt]
{newton@biostat.wisc.edu}}

\markboth%
{T. Vo and others}
{Dimension constraints for testing}
\maketitle
\footnotetext{To whom correspondence should be addressed.}

\begin{abstract}
{For large-scale testing with graph-associated data, we present an empirical Bayes mixture technique to score
local false-discovery rates. Compared to procedures that ignore the graph, the proposed 
GraphMM method gains power in settings where non-null cases form connected subgraphs, and it does so by regularizing 
parameter contrasts between testing units.  Simulations show that GraphMM
controls the false-discovery rate in a variety of settings.   On magnetic resonance imaging
data from a study of brain changes associated with the onset of Alzheimer's disease, GraphMM 
produces substantially greater yield than conventional large-scale testing procedures. }{empirical Bayes, 
graph-respecting partition, GraphMM, image analysis, local false-discovery rate, mixture model}
\end{abstract}

\section{Introduction}
\label{sec1}

Empirical Bayesian methods provide a useful approach to large-scale hypothesis testing in genomics, 
brain-imaging, and other application areas.
Often, these methods are applied relatively late in the data-analysis pipeline, after p-values, test statistics,  or other
summary statistics are computed for each testing unit.  Essentially, the analyst performs univariate testing {\em en masse}, 
with the final unit-specific scores and discoveries dependent upon the chosen  empirical Bayesian method,
which accounts for the collective properties of the separate statistics to gain an advantage
(e.g., \cite{Storey2003}, \cite{Efron_2010}, \cite{Stephens2017}).  These methods are effective but 
may be underpowered in some applied problems
when the underlying effects are relatively weak.  Motivated by tasks in neuroscience, we describe
an empirical Bayesian approach that operates earlier in the data-analysis pipeline and that leverages regularities
achieved by constraining the dimension of the parameter space.  Our approach is restricted to data sets in
which the variables constitute nodes of a known, undirected graph, which we use to guide regularization.   We report
simulation and empirical studies with structural magnetic resonance imaging to demonstrate the striking 
operating characteristics of  the new methodology.  We conjecture that power is gained  for graph-associated data
 by moving upstream in the data reduction process  and by recognizing low complexity parameter states.

\begin{figure}[!p]
\center
\includegraphics[width=0.8\textwidth]{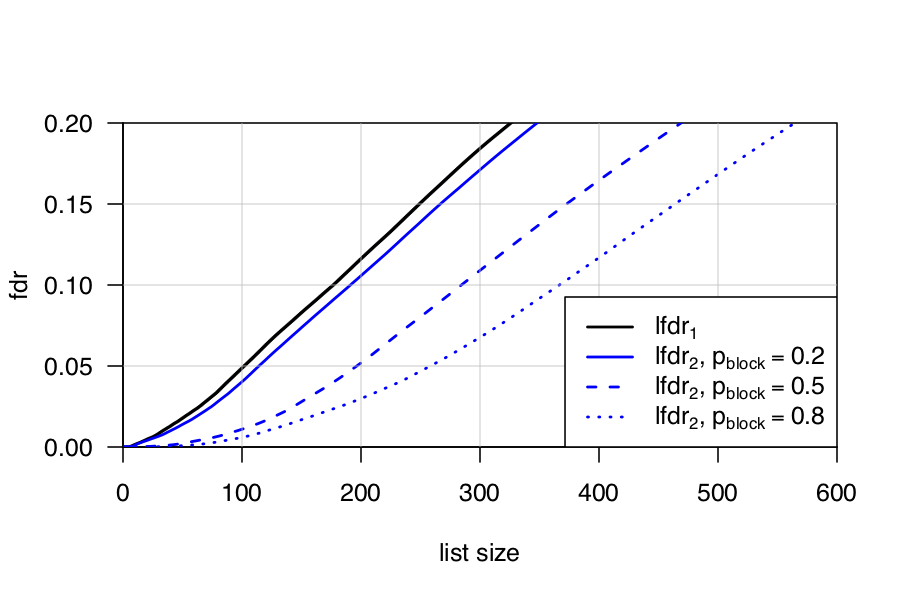}
\caption{False discovery rate (vertical) as a function of list size (horizontal)
 for various testing procedures. ${\rm lfdr}_1$ refers to the procedure
to list the unit if the local false discovery rate
 $P(\mu_{X_1}=\mu_{Y_1}|X_1,Y_1)$ is sufficiently small (black).  Blue
lines refer to the operating characteristics when using ${\rm lfdr}_2$ which
is $P(\mu_{X_1}=\mu_{Y_1}| X_1, X_2, Y_1, Y_2)$, for various probabilities
$p_{\rm block}$ that the two units share parameters.  By accounting for blocking,
we benefit through increased yield for a given false-discovery-rate.
  \label{toy} }
\end{figure}

The following toy problem illustrates in a highly simplified setting
  the phenomenon we  leverage
for improved power.   Suppose we have two sampling conditions, and
two variables measured in each condition,
say $X_1$ and $X_2$ in the first condition and $Y_1$ and $Y_2$ in the second.  We
aim to test the null hypothesis that $X_1$ and $Y_1$ have the same expected value; say
$H_0: \mu_{X_1} = \mu_{Y_1}$.  Conditional upon target values $\mu_{X_1}$, $\mu_{Y_1}$
and  nuisance mean values $\mu_{X_2}$ and $\mu_{Y_2}$, the four observations
are mutually independent, with normal distributions and some constant, known variance
$\sigma^2$.   We further imagine that these four variables are part of a larger system,
throughout which the distinct expected values themselves fluctuate, say according
to a standard normal distribution.   Within this structure, a test of $H_0$ may
be based upon the local false-discovery rate  
\begin{eqnarray*}
{\rm lfdr}_1 = 
 P(H_0|X_1,Y_1) = \frac{p_0 f(X_1,Y_1)}{p_0 f(X_1, Y_1) + (1-p_0) g(X_1) g(Y_1)}
\end{eqnarray*}
where we are mixing discretely over null (with probablity $p_0$) and non-null
cases.  Notice in this setting the across-system variation in expected values
may be handled analytically and integrated out; thus in this predictive
distribution $g$ is the density of 
a mean~0 normal distribution with variance $1+\sigma^2$;
 and $f$ is the bivariate normal density with margins $g$ and
with correlation  $1/(1+\sigma^2)$ between $X_1$ and $Y_1$.  In considering
 data $X_2$ and $Y_2$ on the second variable, it may be useful to suppose
that the expected values here are no different from their counterparts on the first
variable. We say the variables are blocked if both $\mu_{X_1} = \mu_{X_2}$ and 
 $\mu_{Y_1} = \mu_{Y_2}$, and we consider this a discrete possibility that
occurs with probablity $p_{\rm block}$ throughout the system, independently
of $H_0$.  In the
absence of blocking there is no information in $X_2$ and $Y_2$ that could 
inform the test of $H_0$ (considering the independence assumptions).  
In the presence of blocking, however, data on these second variables are highly relevant.
Treating blocking as random variable across the system, we would score $H_0$ using
the local false-discovery rate ${\rm lfdr}_2 = P( H_0 | X_1, X_2, Y_1, Y_2 )$, 
which requires consideration of a 4-variate normal  and joint discrete mixing
over the blocking and null states for full evaluation.
Fig.~\ref{toy} shows the result of simulating
a system with $10^4$ variable pairs, where the marginal null frequency $p_0=0.8$,
$\sigma^2 = 1/2$, and the blocking rate $p_{\rm block}$ varies over three 
possibilities.   Shown is the false-discovery rate of the list formed by ranking
instances by either ${\rm lfdr}_1$ or ${\rm lfdr}_2$.   The finding in this
toy problem is that power for detecting differences between $\mu_{X_1}$ and $\mu_{Y_1}$
 increases by accounting for the blocking, since the list of discovered non-null
cases by ${\rm lfdr}_2$ is larger for a given false-discovery rate than the
list constructed using  ${\rm lfdr}_1$. 
In other words, when the dimension of the parameter space is constrained,
more data become relevant to the test of $H_0$ and power increases.

Our interest in large-scale testing arises from
work with structural magnetic resonance imaging (sMRI) data measured in studies of brain structure and function,
as part of the Alzheimer's Disease Neuroimaging Initiative (ADNI-2) (\cite{ADNI}). 
sMRI provides a detailed view of brain atrophy and has become an integral to the clinical assessment of patients suspected 
to have Alzheimer's disease (AD) (e.g., \cite{Vemuri2010}, \cite{Moller2013}).  In studies to understand 
disease onset, a central task  has been to identify brain regions that exhibit 
 statistically significant differences between various clinical groups, while accounting for technical and
biological sources of variation affecting sMRI scans.
Existing work in large-scale testing for neuroimaging has considered thresholds on voxel-wise test statistics to control a specified false positive rate and maintain testing power (\cite{Nichols2012}). Two widely used approaches
are family-wise error control using random field theory (e.g., \cite{Worsley2004}) and
false-discovery rate control using Benjamin-Hochberg procedure (e.g., \cite{Genovese2002}). The former is based on additional assumptions about the spatial smoothness of the MRI signal which may be problematic (\cite{Eklund2016}).
Both parametric and nonparametric voxel-wise tests are available in convenient neuroimaging software systems 
(\cite{SPM}, \cite{SnPM}).  Recently
\cite{Scott18} presented an FDR tool that processes unit-specific test statistics in a way to spatially smooth
the estimated prior proportions.
As the clinical questions of interest move towards identifying early signs of AD, the changes in average brain profiles between conditions invariably become more subtle and increasingly hard to detect; the result is that very few voxels or brain regions may be detected as significantly different by standard methods.

Making a practical tool from the power boosting idea in  Fig. 1 requires that a number of modeling
and computational issues be resolved. Others have recognized the
potential, and have designed computationally intensive Bayesian approaches  based on Markov chain Monte Carlo
(\cite{Do2005}, \cite{Dahl_Newton_2007}, \cite{Dahl2008}, \cite{Dahl2009}).  We seek simpler methodology that may be
more readily adapted in various applications.  In many contexts
data may be organized at nodes of an undirected graph, which will provide a basis 
for generalizing the concept of blocking using graph-respecting partitions that have a regularizing effect.  
Having replicate observations per group is a basic 
aspect of the data structure, but we must also account for statistical dependence among
variables for effective methodology.  In the proposed formulation 
we avoid the product-partition assumption  that would
 greatly simplify computations but at the
expense of model validity and robustness; we gain numerical efficiency and avoid posterior Markov
chain Monte Carlo through a graph-localization of the mixture computations.
  The resulting tool we call
\verb+GraphMM+, for graph-based mixture model.  It is deployed as a freely available open-source
R package available at \verb+https://github.com/tienv/GraphMM/+. We investigate its
properties using a variety of synthetic-data scenarios, and we also apply it to identify
statistically significant changes in brain structure associated with the onset of mild
cognitive impairment.   Details not found in the following sections are included in Supplementary Material.

\section{Methods}

\subsection{Data structure and inference problem} \label{generalsetting}

\begin{figure}[!p]
\center
\includegraphics[width=0.8\textwidth]{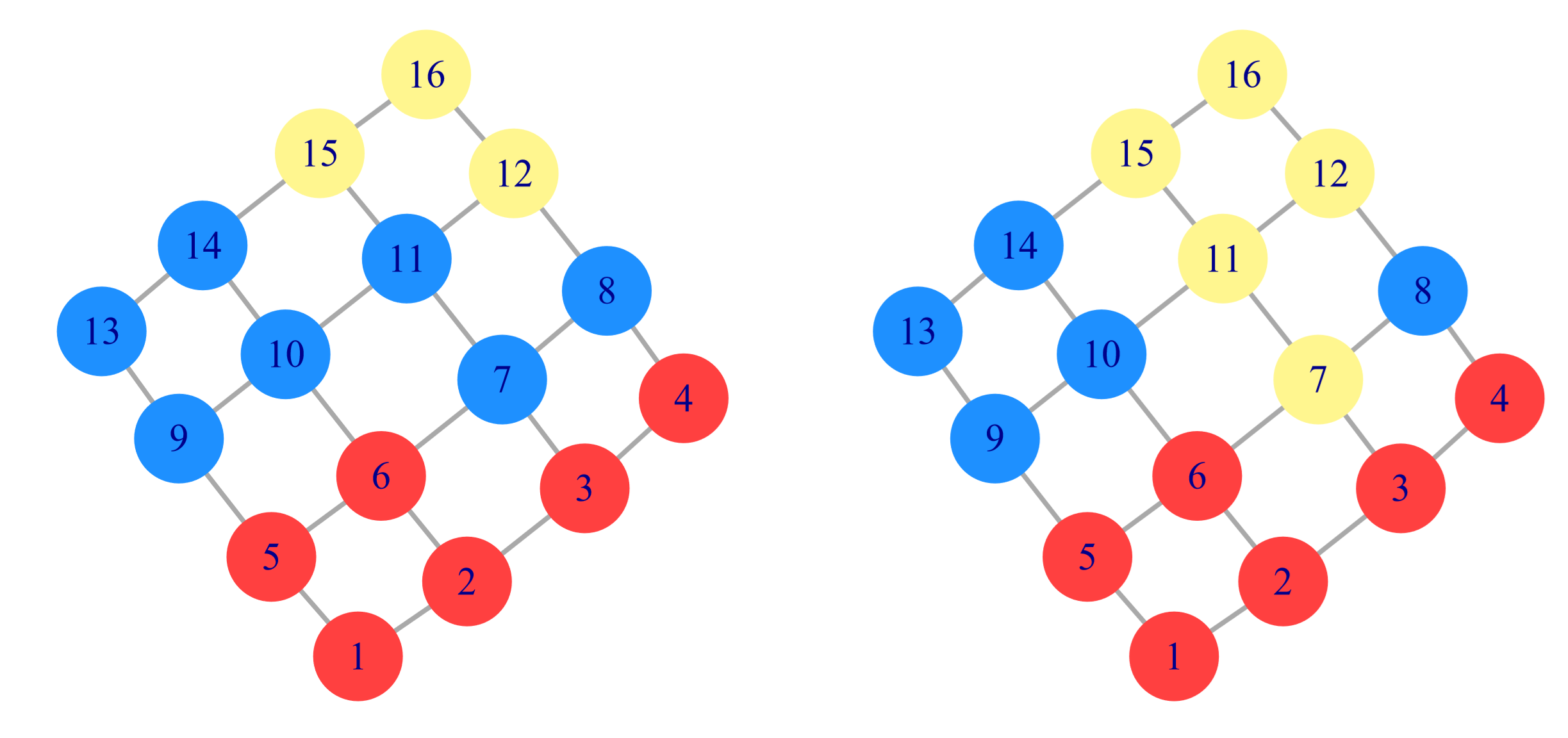}
\caption{Examples of partitions on a graph. Different colors represent different blocks. The partition on the left is graph-respecting while the one on the right is not (the blue block induces a subgraph with two
components).} \label{ExampleGraphPartition}
\end{figure}

Let $G=(V,E)$ denote a simple, connected, undirected graph with vertex set $V = \{1, 2, ..., N\}$ and edge set $E$,  and consider partitions of $V$, such as 
$\Psi = \{ b_1, ..., b_K\}$;  that is, blocks (also called clusters)  $b_k$ constitute
non-empty  disjoint subsets of $V$ for which $\cup_{k=1}^K b_k = V$.  In the application 
in Section 3.2, vertices correspond to voxels at which brain-image data are measured, edges connect 
spatially neighboring voxels, and the partition conveys a  dimension-reducing constraint.  
The framework is quite general and includes, for example, interesting 
problems from genomics and molecular biology.
Recall that for any subset $b \subset V$, the induced subgraph $G_b = (b, E_b)$, where $E_b$ contains all
edges $e=(v_1,v_2)$ for which $e \in E$ and $v_1, v_2 \in b$.   For use in constraining a parameter
space, we introduce the following property:
\begin{property}[Graph respecting partition]
A partition $\Psi$ respects $G$, or 
$\Psi$ is graph-respecting, if for all $b_k \in \Psi$,  the induced graph $G_{b_k}$  is connected.
\end{property}
Fig.~ \ref{ExampleGraphPartition} presents a simple illustration; a spanning-tree representation turns out
to be useful in computations (Supplementary Material).
It becomes relevant 
to statistical modeling that the size of the set of graph-respecting
partitions, though large, still is substantially
smaller than the set of all partitions as the graph itself becomes less complex.
For example
there are 21147 partitions of $N=9$ objects (the 9th Bell number), but if these 9 objects are 
arranged as vertices of  a regular $3 \times 3$ lattice graph, then there are only 1434 graph-respecting
partitions. 

In our setting, the graph $G$ serves as a known object that provides structure to a data set  being analyzed
for the purpose of a two-group comparison. This is in contrast, for example, to graphical-modeling
 settings where the possibly unknown  graph holds the dependency patterns of the joint distribution. 
We write the two-group data as 
$\boldsymbol{X} = (X_{v,m})$ and  $\boldsymbol{Y} = (Y_{v,r})$, where $v \in V$, $ m=1,\dots, M_X $ 
and $ r=1,\dots, M_Y$.  
Here $M_X$ and $M_Y$ denote the numbers of replicate samples in both groups. 
In Section~3.2, for example, $m$ indexes the brain of a normal control subject and $r$ indexes the brain of 
a subject with mild cognitive impairment.  
For convenience, let  $\boldsymbol{X_{m}}=( X_{v,m},\; v \in V )$ and 
$\boldsymbol{Y_{r}}=(Y_{v,r},\; v \in V )$ denote the across-graph 
samples on subjects $m$ and $r$,  which we treat as identically distributed within group and
  mutually independent over $m$ and $r$ 
owing to the two-group, unpaired experimental design.

Our methodology tests for changes between the two groups in the expected-value vectors:
$\boldsymbol{\mu_X} = E( \boldsymbol{X_{m}} ) = (\mu_{X_1}, \dots, \mu_{X_N}) $ 
 and $\boldsymbol{\mu_Y} = E( \boldsymbol{Y_{r}} ) = (\mu_{Y_1}, \dots, \mu_{Y_N})$. Specifically,
we aim to test, for any vertex $v \in V$, 
$H_{0, v}: \mu_{X_v} = \mu_{Y_v}$  versus $H_{1, v}: \mu_{X_v} \neq \mu_{Y_v}$.
We seek to gain statistical power over contemporary testing procedures by imposing a dimension
constraint on the expected values.  Although it is not required to be known or even 
estimated, we suppose there exists a graph-respecting partition $\Psi=\{b_k\}$ that constrains the expected 
values:
\begin{equation} \label{meanclusX}
\begin{cases}
\mu_{X_v} = \mu_{X_u}     \quad \quad  \textrm{if for some $k$, both}\, v, u \in b_k \\
 \mu_{X_v} \neq \mu_{X_u} \quad \quad  \textrm{if} \, v, u \textrm{ belong to different   blocks } 
\end{cases} 
\end{equation}
\begin{equation*} 
\begin{cases}
 \mu_{Y_v} = \mu_{Y_u}     \quad \quad  \textrm{if for some $k$, both}\,  v, u \in b_k \\
 \mu_{Y_v} \neq \mu_{Y_u}  \quad \quad  \textrm{if} \,  v, u \textrm{ belong to different blocks } 
\end{cases} 
\end{equation*}
 All vertices $v$ in  block $b_k$ have a common mean in the first group, say 
$\varphi_k$, and a common mean $\nu_k$ in the second group.
The contrast on test, then, is $\delta_k = \nu_k - \varphi_k$; together with $\Psi$, the binary vector 
$\boldsymbol{\Delta} = (\Delta_1, \dots , \Delta_K )$ holding  indicators $\Delta_k = 1[ \delta_k \neq 0 ]$
is equivalent to knowing whether or not $H_{0,v}$ is true for each vertex $v$.
When data are consistent with a partition $\Psi$ in which the number of blocks $K$ is  small compared to 
the number of vertices $N$, then it may be possible to leverage this reduced parameter-space 
complexity for the benefit of hypothesis-testing power.

\subsection{Graph-based Mixture Model}  \label{datamodel}

\subsubsection{Discrete mixing.} 
We adopt an empirical Bayes, mixture-based  testing approach, which requires that for each vertex we 
compute a local false-discovery rate:
\begin{equation} \label{postprob}
l_v := P(H_{0, v} | \boldsymbol{X}, \boldsymbol{Y}) = 
\sum_{\Psi, \boldsymbol{\Delta}} (1-\Delta_k) \mathds{1}( v \in b_k)P(\boldsymbol{\Delta}, \Psi|\boldsymbol{X}, \boldsymbol{Y}).
\end{equation}
Our list  $\mathcal L$ of discovered (non-null) vertices is $\mathcal{L} = \{ v: l_v \leq c \}$
for some threshold $c$. Conditional on the data, the expected rate of type-I errors within $\mathcal L$
is dominated by the threshold $c$ (\cite{Efron2007},  
 \cite{Newton2004DetectingDG}).          
The sum in~(\ref{postprob}) is over the finite set of pairs of partitions $\Psi$ and block-change indicator
vectors $\boldsymbol{\Delta}$.   
This set is intractably large for even moderate-sized graphs.   We have experimented
with Markov-chain Monte Carlo for general graphs, but present here exact computations in the context of
very small graphs.   Specifically, for each vertex $v$ in the original graph we consider a small
local subgraph in which $v$ is one of the central vertices, and we simply deploy \verb+GraphMM+
on this local graph.  

Summing in~(\ref{postprob}) delivers marginal posterior inference, and  thus a mechanism for
borrowing strength among vertices $v$.   By Bayes's rule,
$P(\boldsymbol{\Delta}, \Psi|\boldsymbol{X}, \boldsymbol{Y})
\propto f(\boldsymbol{X}, \boldsymbol{Y} | \boldsymbol{\Delta}, \Psi ) \, 
 P(\boldsymbol{\Delta}, \Psi )$,
and both the mass $P(\boldsymbol{\Delta}, \Psi )$ and the predictive density
$f(\boldsymbol{X}, \boldsymbol{Y} | \boldsymbol{\Delta}, \Psi )$ need to be specified to compute inference
summaries. Various modeling approaches present themselves. For example, we could reduce data per vertex to
a test statistic (e.g., t-statistic) and model the predictive density nonparametrically,
as in the R package \verb+locFDR+ (see \cite{Efron_2010}).   Alternatively, we could reduce data per vertex
less severely, retaining effect estimates and estimated standard errors, as in adaptive
shrinkage (\cite{Stephens2017}).  
By contrast, we adopt an explicit parametric-model formulation
for the predictive distribution of data given the discrete state $(\Psi, \boldsymbol{\Delta})$.
It restricts the sampling model to be Gaussian, but allows general covariance among vertices
and is not reliant on the product-partition assumption commonly used in partition-based models (\cite{Barry1992}).
For $P(\Psi,\Delta)$,  we specify $P(\Psi) \propto 1$, we
encode
independent and identically distributed  block-specific Bernoulli$(p_0)$ indicators in $P(\Delta|\Psi)$, and
 we use univariate empirical-Bayes techniques to estimate $p_0$.

\subsubsection{Predictive density given discrete structure.}
We take a multivariate Gaussian sampling model:
\begin{eqnarray*}
\boldsymbol{X_{m}} | \boldsymbol{\mu_X}, U, \Psi, \boldsymbol{\Delta}   \sim_{\rm i.i.d.}  \mathcal{N}(\boldsymbol{\mu_X}, U) \quad  m = 1, \dots, M_X,  \quad
\boldsymbol{Y_r} | \boldsymbol{\mu_Y}, W, \Psi, \boldsymbol{\Delta}   \sim_{\rm i.i.d.}   \mathcal{N}(\boldsymbol{\mu_Y}, W) \quad  r = 1, \dots, M_Y.
\end{eqnarray*}
We do not constrain the $N \times N$ covariance matrices $U$ and $W$, though we place a conjugage inverse Wishart prior distribution on them: $ U | \Psi, \boldsymbol{\Delta}, \boldsymbol{\mu_X}, \boldsymbol{\mu_Y}  
 \sim  \mathcal{IW}(A, \textrm{df})$, and
$W | \Psi, \boldsymbol{\Delta}, \boldsymbol{\mu_X}, \boldsymbol{\mu_Y}  
  \sim  \mathcal{IW}(B, \textrm{df})$.
In general there is no simple conjugate reduction for predictive densities owing to the 
less-than-full dimension
of free parameters in $\boldsymbol{\mu_X}$ and $\boldsymbol{\mu_Y}$. On these free parameters
we further specify independent Gaussian priors:
$\varphi_k  \sim  \mathcal{N}\left(\mu_0, \tau^2 \right)$  and, for $\Delta_k \neq 0$,
$\delta_k  \sim  \mathcal{N}\left(\delta_0, \sigma^2\right)$.
 Hyperparameters in \verb+GraphMM+ 
include scalars $\delta_0$, $\mu_0$, $\tau^2$, $\sigma^2$, df, and 
matrices $A$, $B$, which we estimate from data across the whole graph following the 
empirical-Bayes approach.

The above specification induces a joint density
 $f( \boldsymbol{X}, \boldsymbol{Y}, \boldsymbol{\mu_X}, \boldsymbol{\mu_Y}, U, W | \boldsymbol{\Delta}, \Psi)$. For the purpose of hypothesis testing, 
we need to marginalize most variables, since $H_{0,v}$ is equivalent to $\Delta_k = 0$ and
$v \in b_k$ for block $b_k$ in partition $\Psi$, and local false-discovery rates require
marginal posterior probabilities. 
 Integrating out inverse Wishart distributions over the covariance matrices is possible
analytically.  We find:
\begin{eqnarray} \label{lkh}
f(\boldsymbol{X}, \boldsymbol{Y} \,\, |\,\, \boldsymbol{\mu_X}, \boldsymbol{\mu_Y},  \boldsymbol{\Delta}, \Psi) & = & C \frac{|A|^{\frac{\textrm{df}}{2}} |B|^{\frac{\textrm{df}}{2}}}{|\widetilde{A}|^{\frac{\textrm{\textrm{df}}+M_X}{2}} |\widetilde{B}|^{\frac{\textrm{\textrm{df}}+M_Y}{2}}} 
\end{eqnarray}
where 
$$S_1 = \dfrac{1}{M_X-1} \sum_{m=1}^{M_X}(\boldsymbol{X_m} - \overline{\boldsymbol{X}})(\boldsymbol{X_m} - \overline{\boldsymbol{X}})^\tau, \qquad 
S_2 = (\overline{\boldsymbol{X}}-\boldsymbol{\mu_X})(\overline{\boldsymbol{X}}-\boldsymbol{\mu_X})^\tau $$
$$T_1 = \dfrac{1}{M_Y-1} \sum_{r=1}^{M_Y}(\boldsymbol{Y_r} - \overline{\boldsymbol{Y}})(\boldsymbol{Y_r} - \overline{\boldsymbol{Y}})^\tau, \qquad 
T_2 = (\overline{\boldsymbol{Y}}-\boldsymbol{\mu_Y})(\overline{\boldsymbol{Y}} -\boldsymbol{\mu_Y})^\tau $$
$$\widetilde{A} = A + (M_X-1)S_1 + M_X S_2, \qquad 
\widetilde{B} = B + (M_Y-1)T_1 + M_Y T_2 $$
and where $C$ is a normalizing constant.
In the above, $|.|$ denotes matrix determinant, 
$ \overline{\boldsymbol{X}} = \frac{1}{M_X}\sum_{m=1}^{M_X} \boldsymbol{X_m}$, 
$\overline{\boldsymbol{Y}} = \frac{1}{M_Y} \sum_{r=1}^{M_Y} \boldsymbol{Y_r},$ and
$S_1$ and $T_1$ are sample covariance matrices of $\boldsymbol{X}$ and $\boldsymbol{Y}$.  
In~(\ref{lkh}) there is conditional independence
of data from the two conditions given the means but marginal to the unspecified covariance matrices.
We use the Laplace approximation to numerically integrate the freely-varying  means in order 
to obtain the marginal predictive density 
$f(\boldsymbol{X}, \boldsymbol{Y} | \boldsymbol{\Delta}, \Psi)$ (Supplementary Material, Equation 0.2).

By not constraining the sample covariance matrices $U$ and $V$ the \verb+GraphMM+ model
does not adopt a product-partition form. In such, the predictive density would factor over blocks
in the graph-respecting partition, and this would lead to simpler computations.  We found in
preliminary numerical experiments that various data sets are not consistent with this simplified
dependence pattern, and we therefore propose the general form here.    
Per voxel computations remain relatively efficient since we work on small, local graphs.

\subsection{Data-driven simulations} \label{simproc}

Our primary evaluation of \verb+GraphMM+ is through a set of simulations
designed around a motivating data set from the
 Alzheimer's Disease Neuroimaging Initiative 2 (ADNI-2) (Section 2.4).
 Briefly,  we consider structural brain imaging data from a group of $M_X=123$ normal control subjects (group 1) 
and a second group of $M_Y = 148$ subjects suffering from late-stage mild cognitive impairment 
(MCI), a precursor to Alzheimer's disease (AD), and for the simulation we focus on 
 a single coronal slice containing $N=5236$ voxels.
We cluster the empirical mean profiles to generate blocks for
the synthetic expected values, and we use covariance matrices estimated from the replicate brain slices. 
 Three synthetic data sets are generated in each simulation scenario.
The first  three  scenarios address the issue of block size; the next two investigate the role of 
the distribution of condition effects.   To assess robustness,
we also  consider parameter settings where partitions are not
graph respecting, and condition effects are not uniform over blocks.   Further, we deploy two permutation
experiments; the first uses sample label permutation to confirm the control of the false-discovery rate,
and the second uses voxel permutation to confirm that sensitivity drops when we disrupt the spatially
coordiated signal.

When applying \verb+GraphMM+ to each synthetic data set, we estimate hyperparameters for all distributional components
and consider  discoveries as ${\mathcal L}(c) = \{ v: l_v \leq c \}$ for various thresholds $c$.  
We call the controlled FDR the
mean $\sum_v l_v 1[ v \in {\mathcal L}(c) ]/\sum_v 1[v \in {\mathcal L}(c)]$, as this is the conditional expected rate
of type-1 errors on the list, given data (and computable from data). 
 We know the null status in each synthetic case, and so we also
call the empirical FDR to be that rate counting latent null indicators; likewise the true positive rate counts 
the non-null indicators.  We compare \verb+GraphMM+ to several contemporary testing methods, including 
Benjamini-Hochberg correction (\texttt{BH adj}), \cite{BH1995}, local FDR, (\texttt{locfdr}) \cite{Efron2007}, 
and  $q$-value (\texttt{qvalue}), \cite{Storey2003}, that are
 all applied to voxel-specific t-tests.  We also compare results to adaptive shrinkage, \cite{Stephens2017},
both the local FDR statistic (\texttt{ash\_lfdr}) and the $q$-value (\texttt{ash\_qval}).
These methods do not leverage the graphical nature of the data, and all work on summaries of  voxel-specific tests; 
summaries may be p-values (for BH and q-value), or t-statistics (for locFDR),
or effect estimates and estimated standard errors (for ASH).  

\subsection{Brain MRI study}
\label{sec:brain-mr}

Gray matter tissue probability maps derived from the co-registered T1-weighted magnetic
resonance imaging (MRI) data  were pre-processed using
the voxel-based morphometry (VBM) toolbox in Statistical Parametric Mapping software (SPM, http://www.fil.ion.ucl.ac.uk/spm).
The ADNI-2 data includes 3D brain images of $148$ cognitively normal control subjects
 and $123$ subjects with late-stage mild cognitive impairment (MCI).
Prior to registration to a common template,
standard artifact removal and other corrections were performed, as described in~\cite{ithapu2015imaging}.
We filtered voxels having very low marginal standard deviation~\citep{IndFilter}, leaving $464441$ voxels,
and then converted the data to rank-based normal scores.

\section{Results}

\subsection{Operating characteristics of {\em GraphMM}}

\begin{figure}[p]
\centering
\begin{tabular}[t]{lll}
\parbox[c]{1em}{\includegraphics[width=0.42\textwidth]{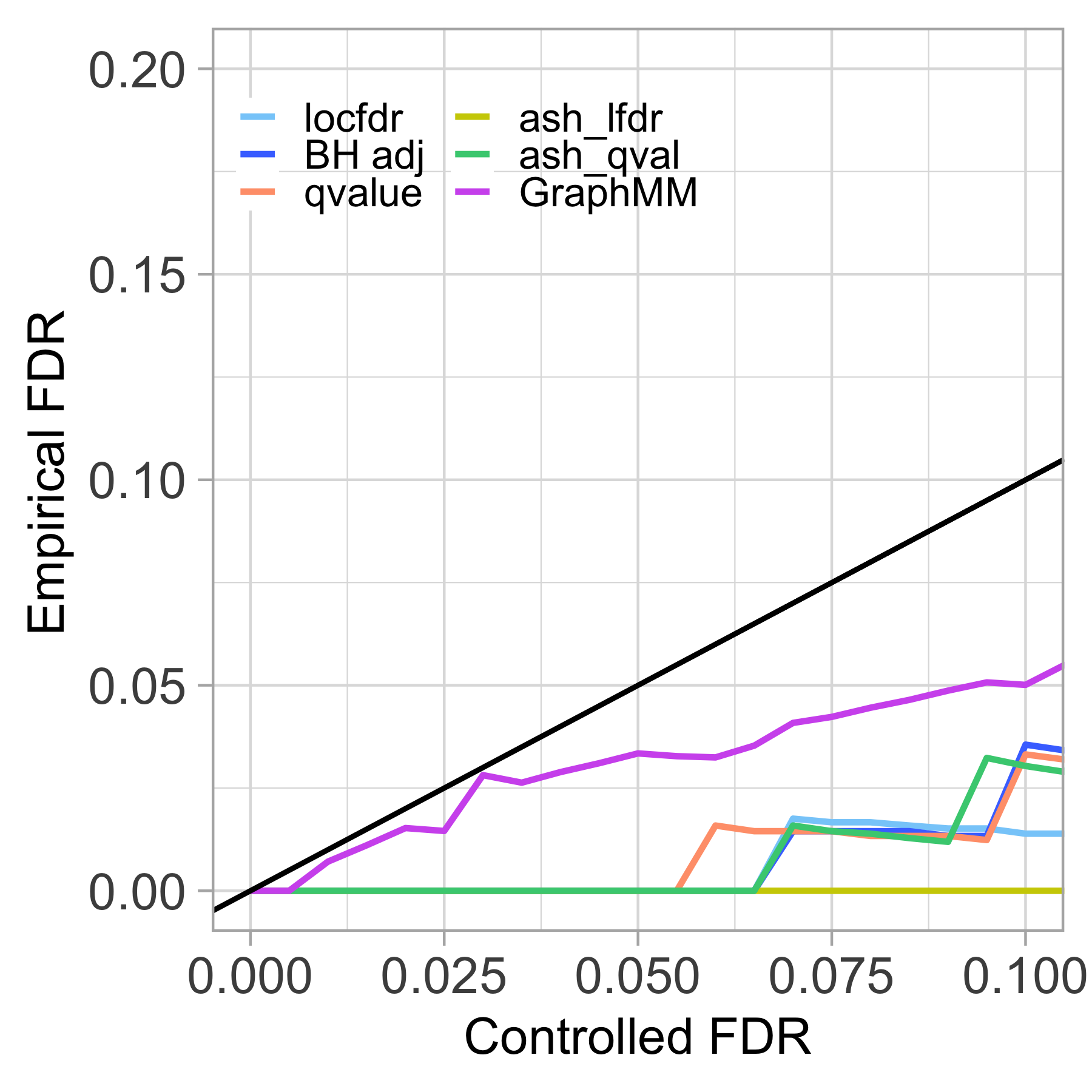}} &
\parbox[c]{1em}{\includegraphics[width=0.42\textwidth]{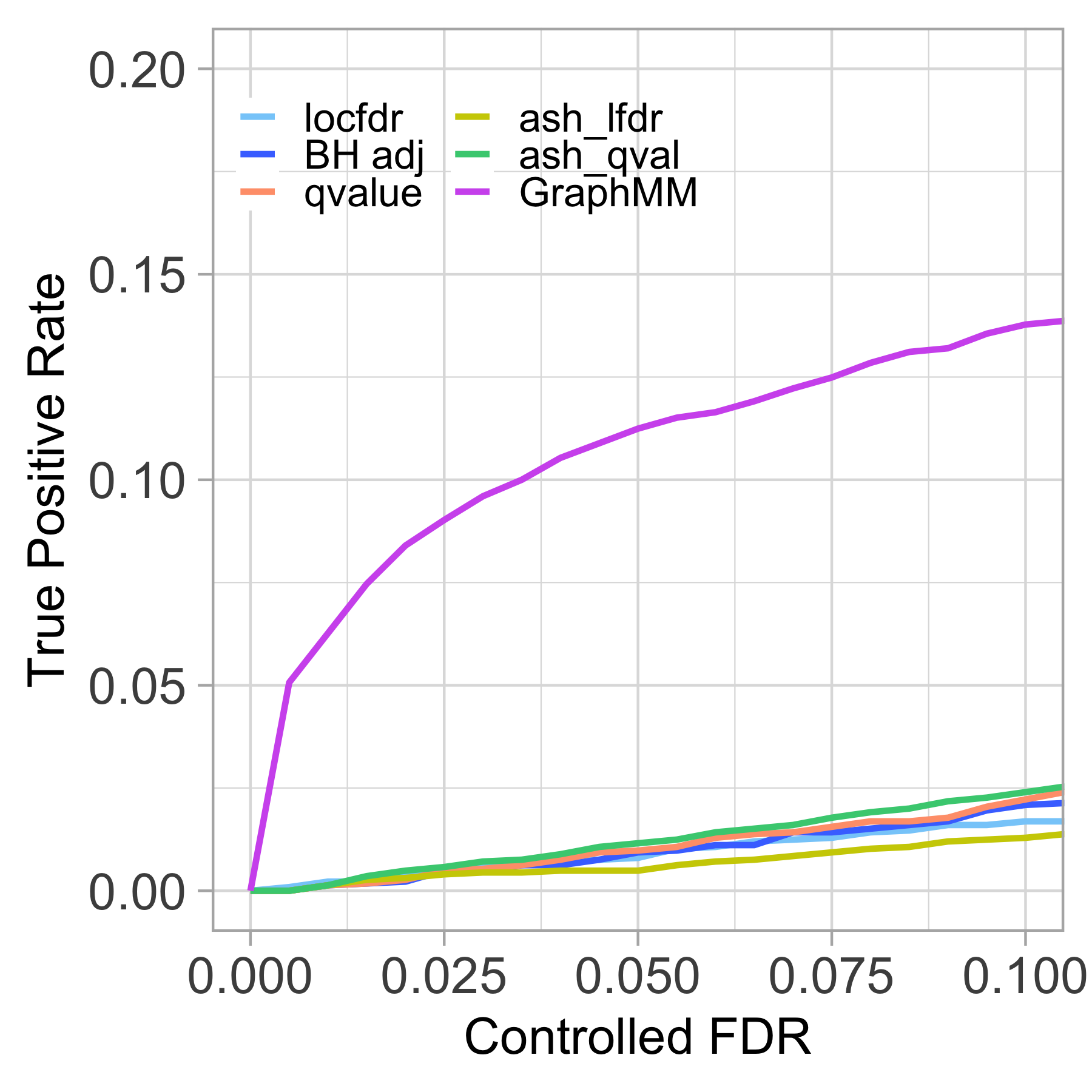}} &
   \begin{tabular}{l}
   Scenario 1 \\
   12-14 vx/block  \\
  \end{tabular} \\
\raisebox{-.5\height}{\includegraphics[width=0.42\textwidth]{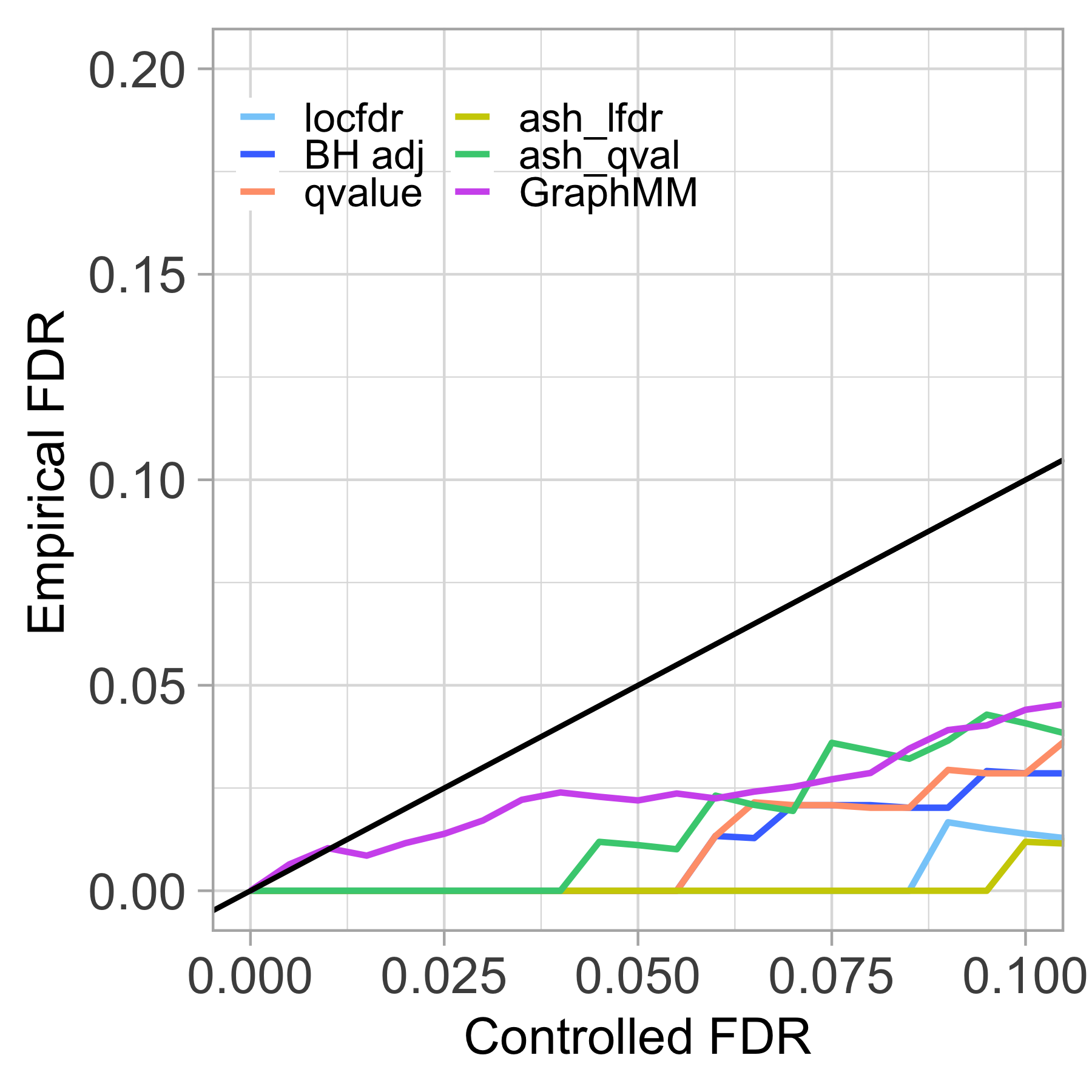}} &
\raisebox{-.5\height}{\includegraphics[width=0.42\textwidth]{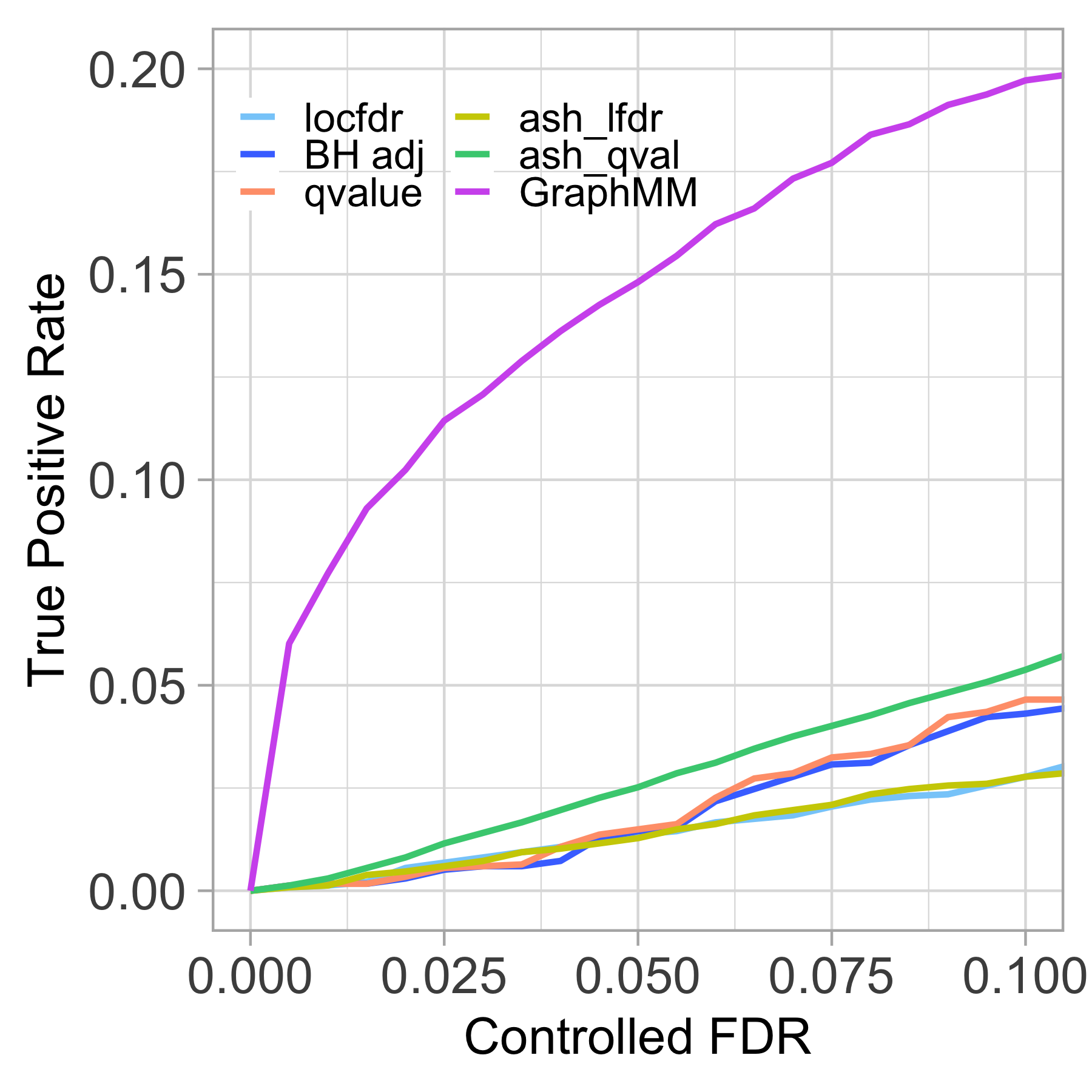}}
 &
 \begin{tabular}{ll}
  Scenario 2 \\
  2-5 vx/block \\
 \end{tabular}    \\
\raisebox{-.5\height}{\includegraphics[width=0.42\textwidth]{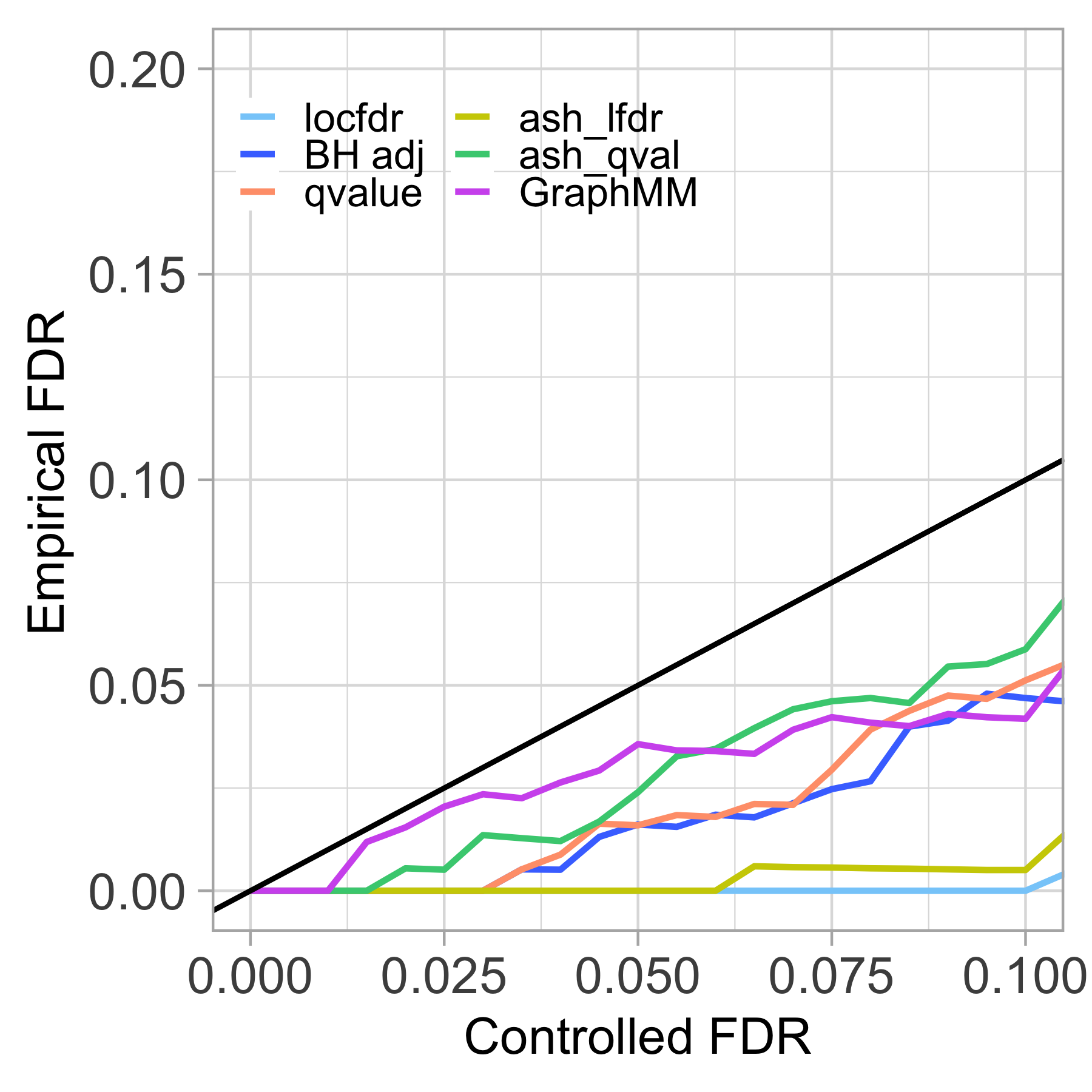}} &
\raisebox{-.5\height}{\includegraphics[width=0.42\textwidth]{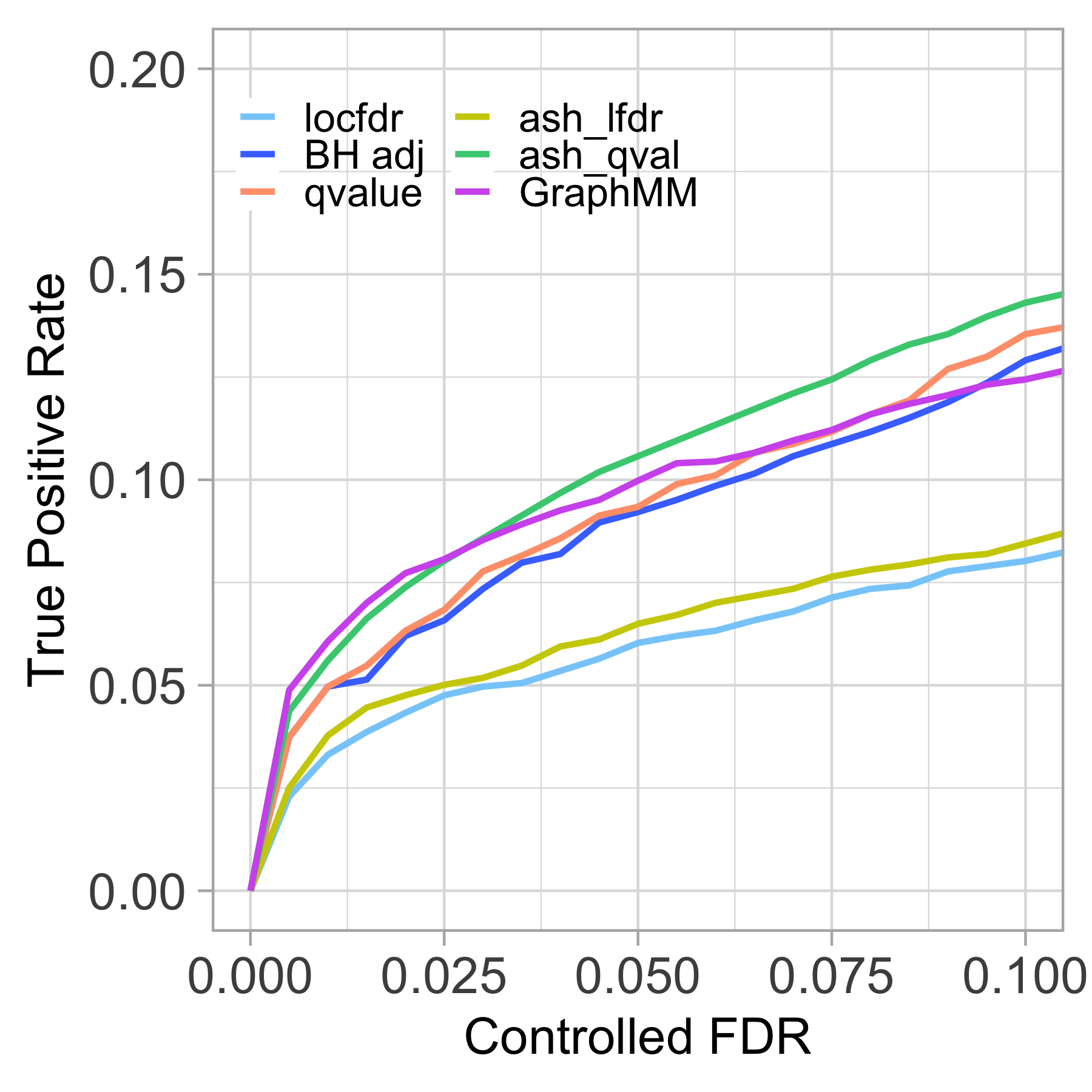}}
  &  \begin{tabular}{l}
      Scenario 3 \\
      1 vx/block \\
     \end{tabular} \\
\end{tabular}
\caption{Operating characteristics on synthetic data.
Rows correspond to simulation scenarios (1=top, large blocks; 2=middle, small blocks;
 3=bottom, tiny blocks). On the left we
 compare the empirical FDR with the target controlling FDR.  Dominance by the diagonal (black)
confirms that all methods are controlling FDR at the target rates.   The right panels
 show how well different methods identify voxels that are truly different between the two
 groups. Substantial power gains are evident by GraphMM. Supplementary Table~S1 provides simulation details.} \label{S123}
\end{figure}

 Synthetic data sets mimic the structure
and empirical characteristics of the brain MRI study data.
The first three synthetic-data scenarios consider a single MRI brain slice
 measured on replicates from two conditions, with characteristics approximately matching
 the characteristics of observed data  (Supplementary Material, Table~S2).  These scenarios vary the 
underlying size distribution of blocks, but follow the \verb+GraphMM+ model in
the sense that the underlying signal has graph-respecting partitions, and other conditions such as block-level shifts between
conditions and multivariate Gaussian errors are satisfied.
The left panel of Fig.~\ref{S123} shows that all methods on test are able to control 
the false-discovery rate.   All methods display sensitivity for the signals, though \verb+GraphMM+
demonstrates superior power in the first two cases where blocks extend beyond the individual
voxel.   The high sensitivity in Scenario~2 may reflect that the prior distribution of block sizes
used in the local \verb+GraphMM+ more closely matches the generative situation.  Notably, even
when this block-size distribution is not aligned with the \verb+GraphMM+ prior, we do not see
an inflation of the false-discovery rate.

\begin{figure}[p]
\centering
\begin{tabular}[t]{lll}
\parbox[c]{1em}{\includegraphics[width=0.42\textwidth]{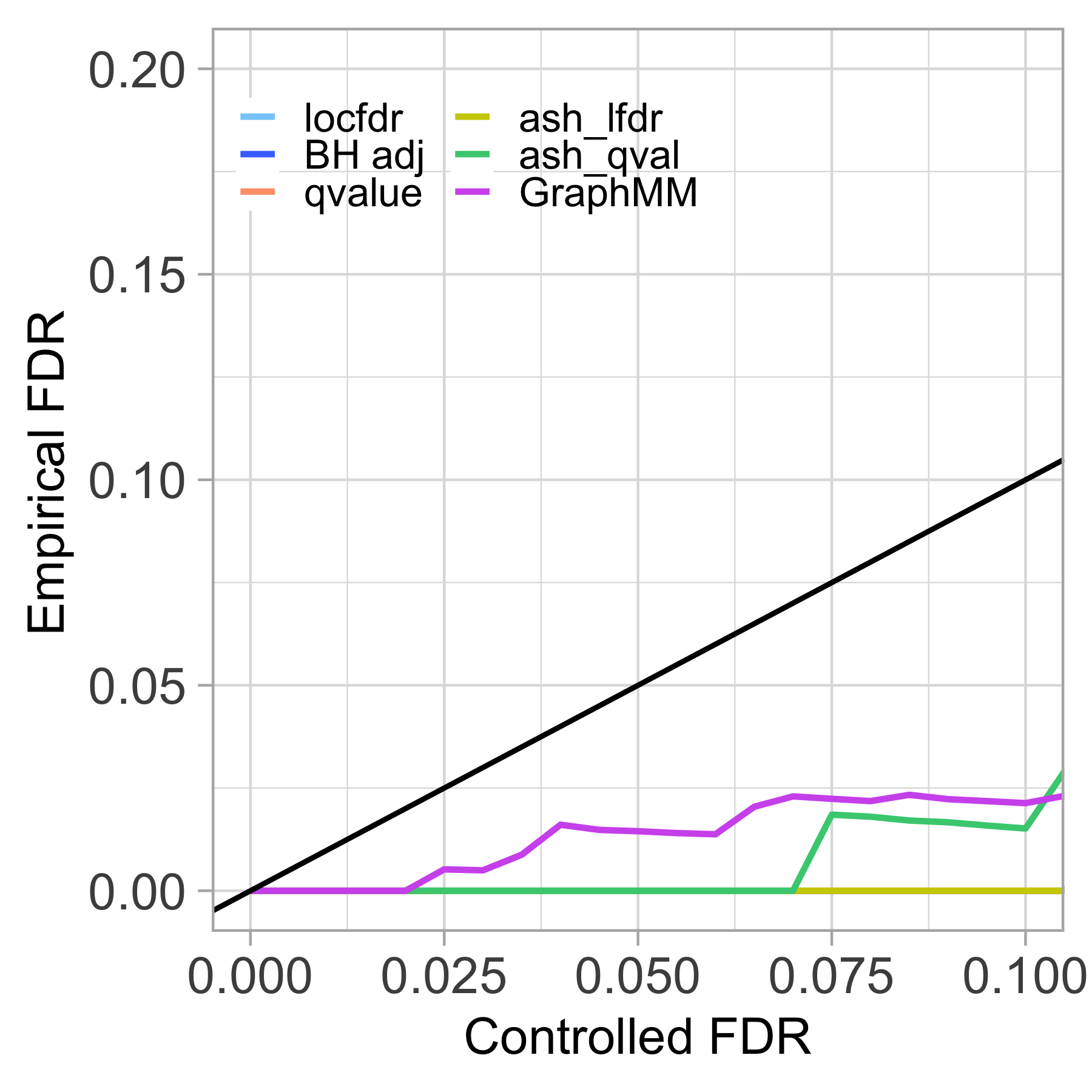}} &
\parbox[c]{1em}{\includegraphics[width=0.42\textwidth]{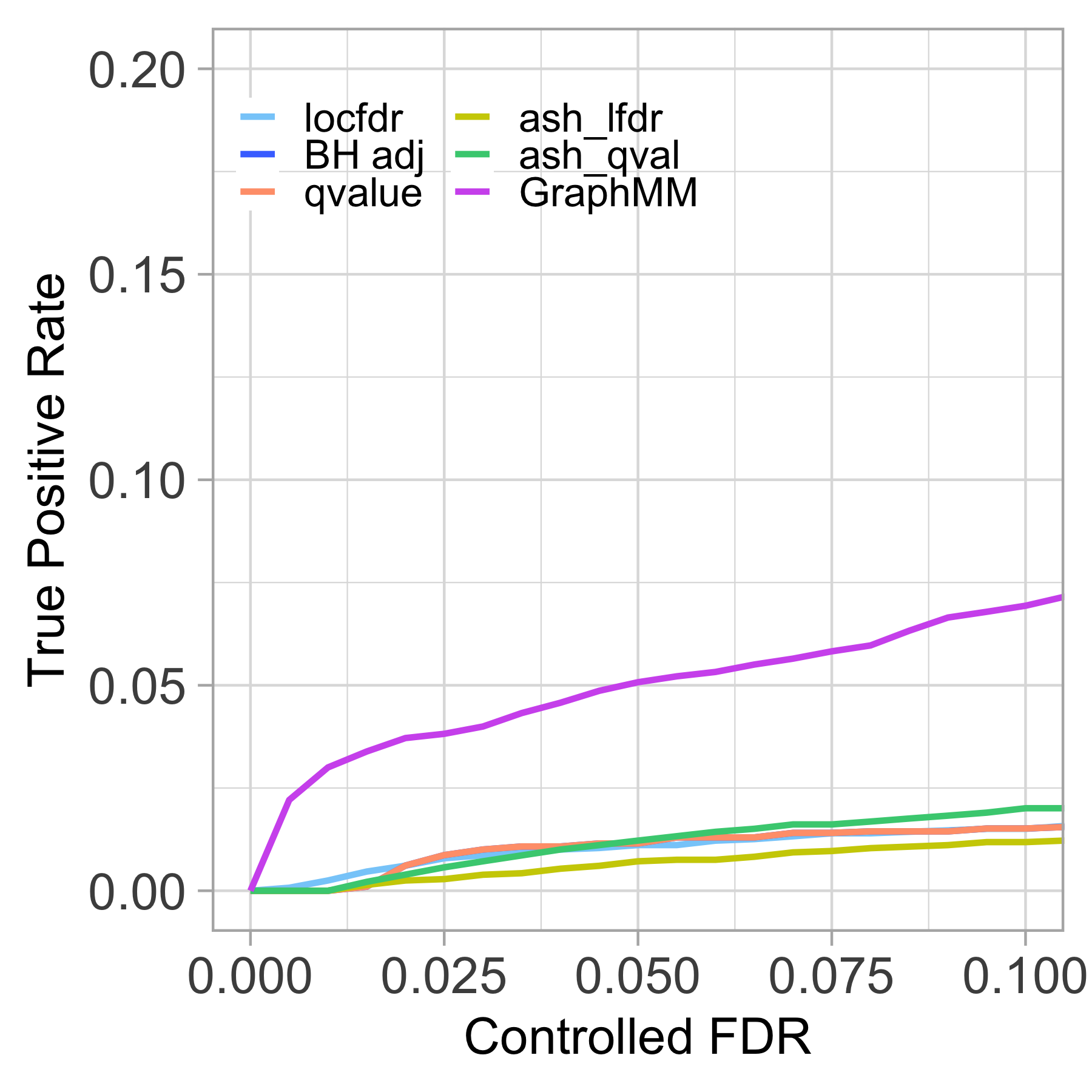}} &
  \begin{tabular}{ll}
   Scenario 4 \\
   low frequency \\
   large shifts  \\
  \end{tabular} \\
\raisebox{-.5\height}{\includegraphics[width=0.42\textwidth]{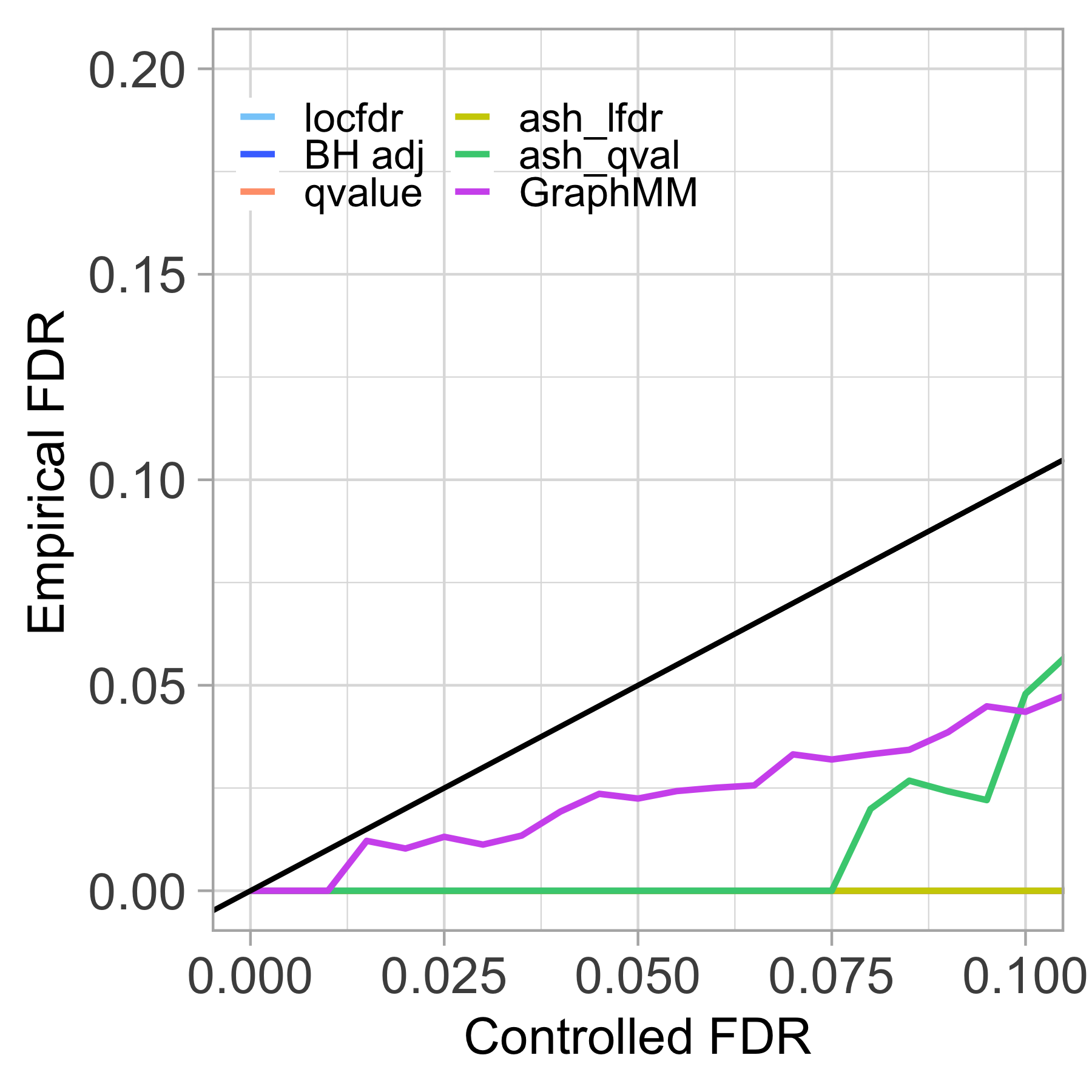}} &
\raisebox{-.5\height}{\includegraphics[width=0.42\textwidth]{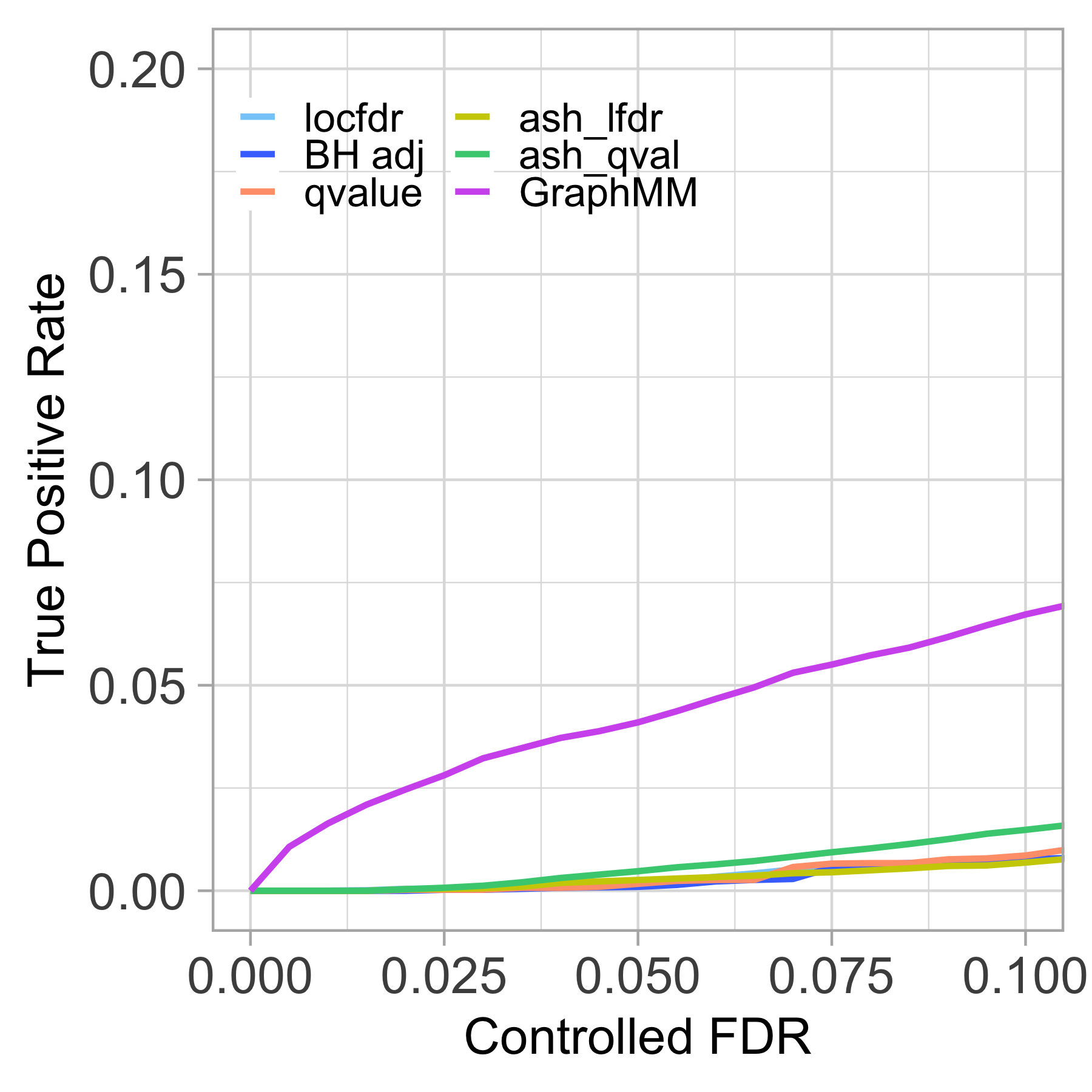}}
 &  \begin{tabular}{ll}
     Scenario 5 \\
     high frequency \\
     small shifts \\
    \end{tabular}    \\
\end{tabular}
\caption{Operating characteristics on synthetic data which explore different distributions of
 effects (Scenarios 4 and 5). Supplementary
Table~S2 has details.  }\label{S45}
\end{figure}

\begin{figure}[p]
\begin{tabular}{cc}
\includegraphics[width=.42\textwidth]{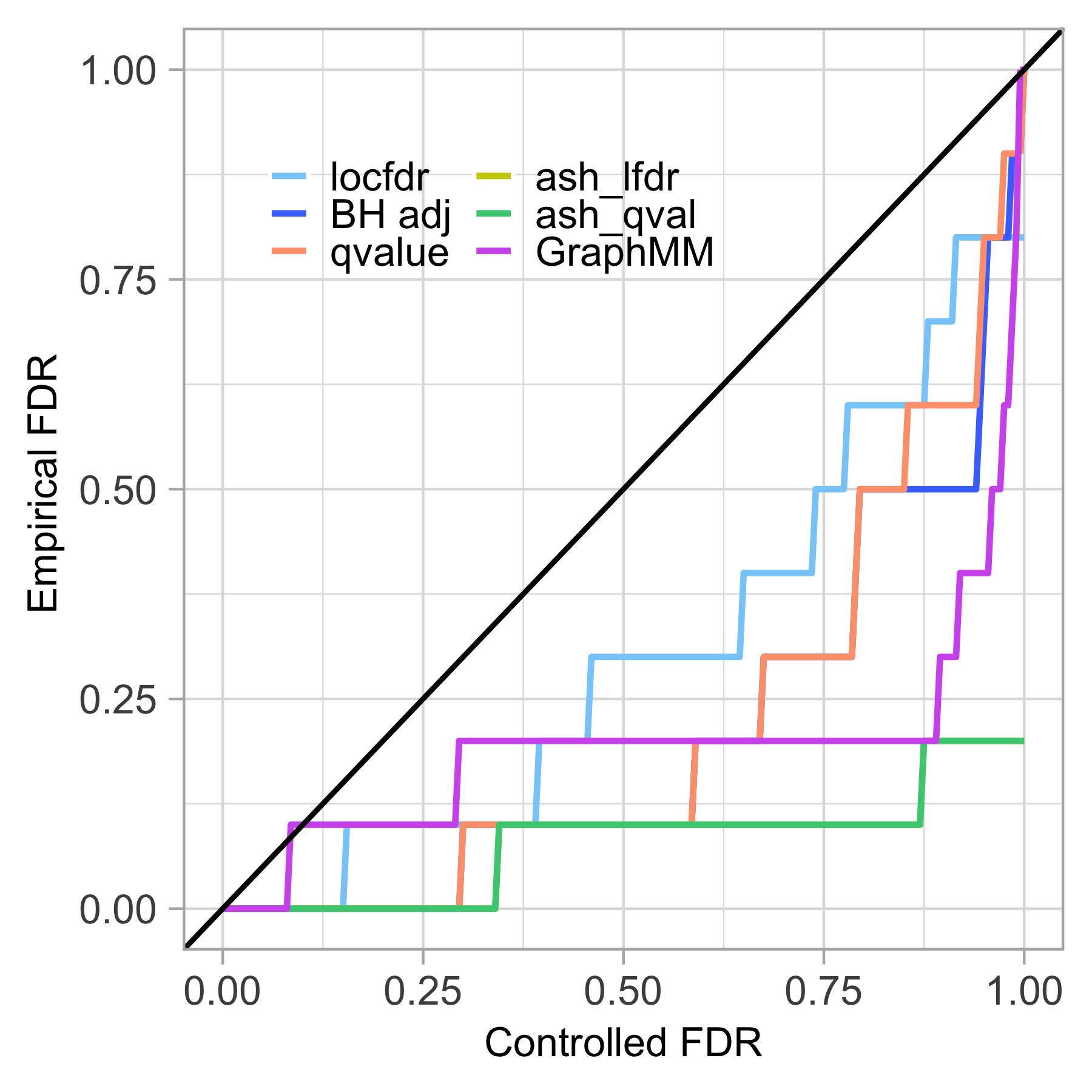} &
\includegraphics[width=.42\textwidth]{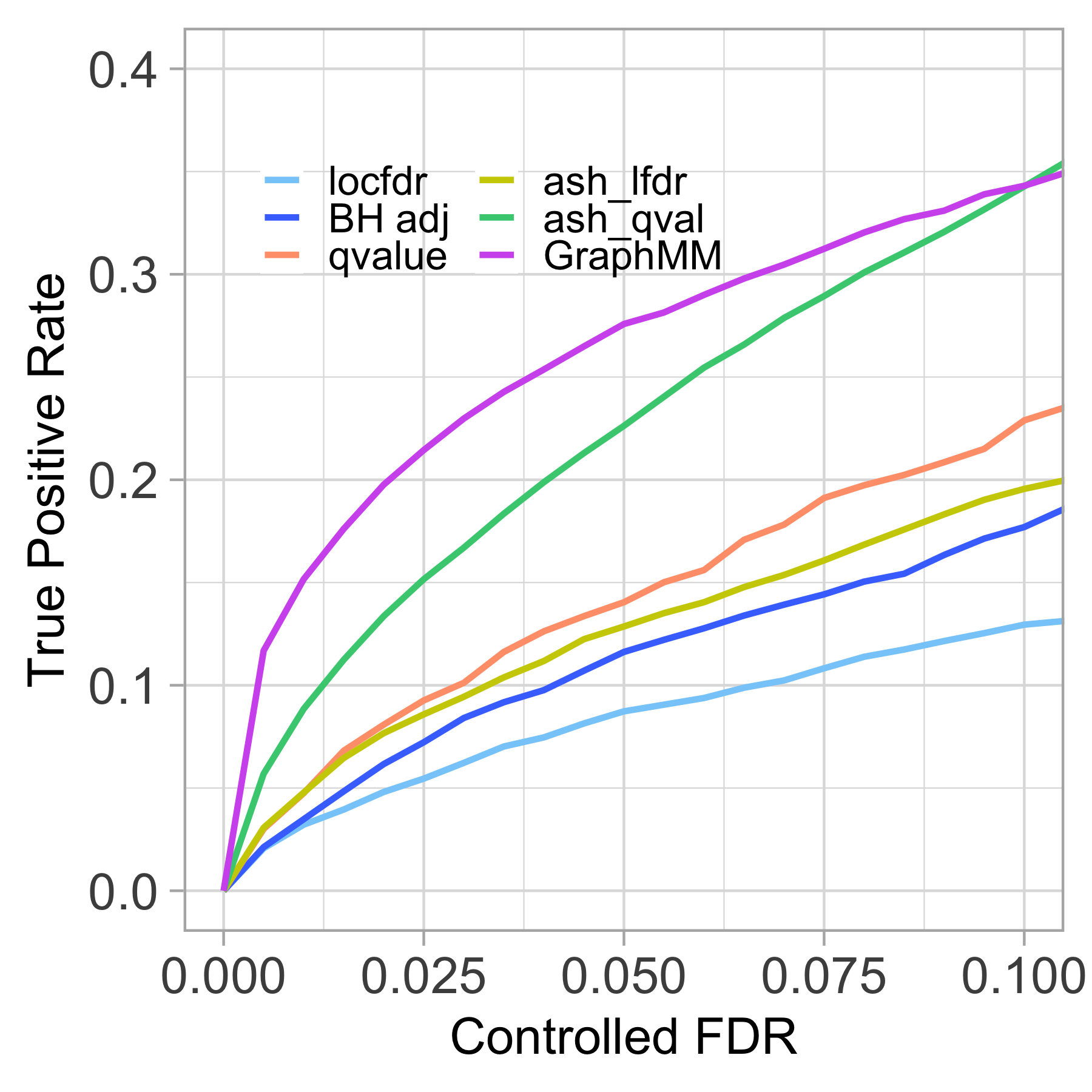} \\
\end{tabular}
\caption{Robustness to graph-respecting assumption. FDR (left) and sensitivity (right)
in a case where latent partitions are not graph respecting, but have similarity 0.48 (Rand index).
} \label{robust}
\end{figure}

Scenarios~4 and~5  are similar to the first cases, however they explore different forms
of signals between the two groups; both have an average block size of 4 voxels, but in one
case the changed block effects are fewer, relatively strong and in the
other case they are more frequent, and relatively weaker (Supplementary Material, Table~S3).  
 In both regimes, \verb+GraphMM+
retains its control of FDR and exhibits good sensitivity compared to other methods (Fig.~\ref{S45}).

\verb+GraphMM+ is designed for the case where partition blocks are graph respecting and the 
changes between conditions affect entire blocks.   Our next numerical experiment checks the
robustness of \verb+GraphMM+ when this partition/change structure is violated 
Fig.~\ref{robust} shows that  \verb+GraphMM+ continues to  control FDR and also 
retains a sensitivity advantage even when its underlying model is not fully correct.

\begin{figure}[p]
\begin{tabular}{cc}
{\includegraphics[width=0.45\textwidth]{FDRPlot.png} } &
{\includegraphics[width=0.45\textwidth]{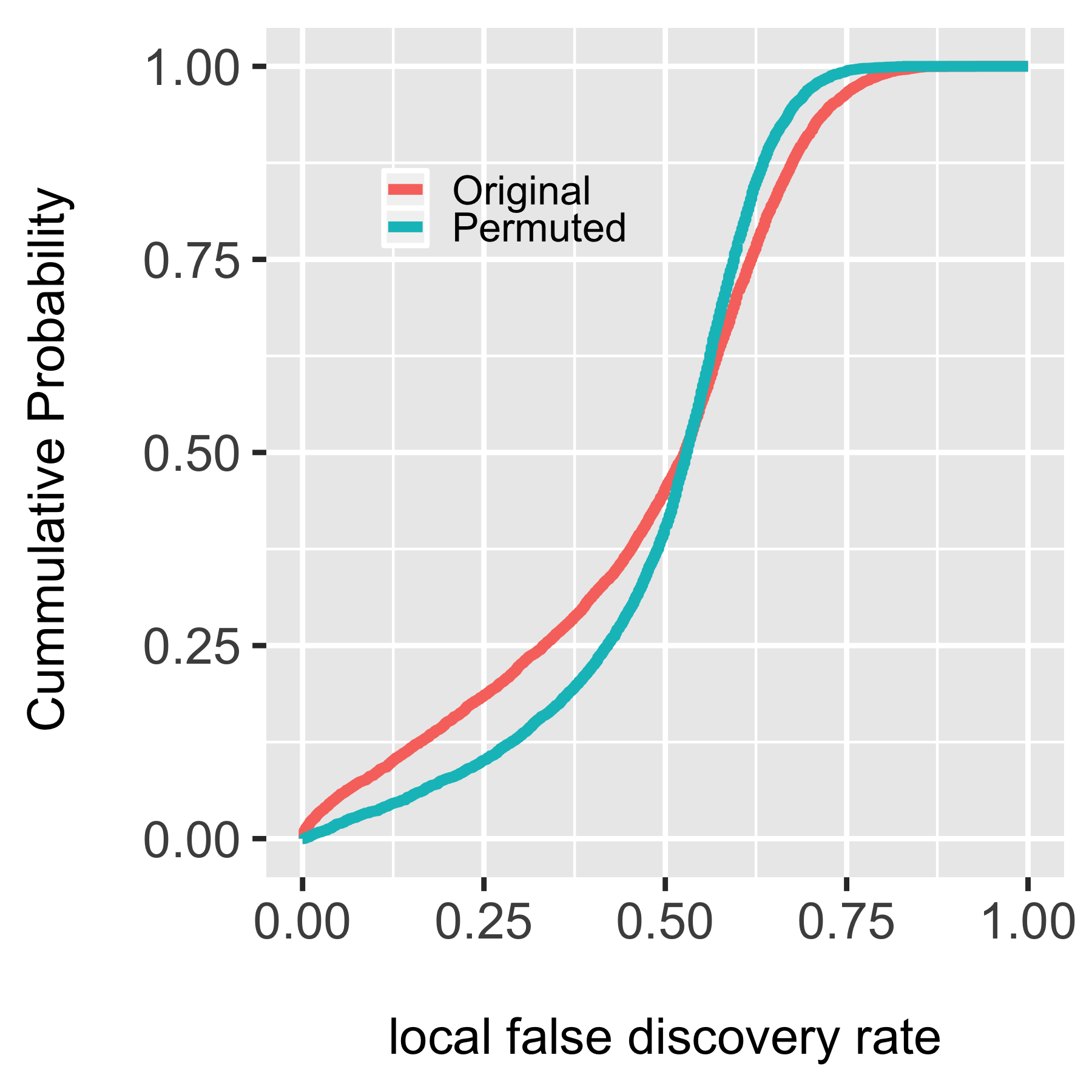} } \\
 sample label permutation &  voxel permutation \\
\end{tabular}
\caption{Permutation experiments: The sample-label-permutation experiment confirms that GraphMM
controls FDR in this no-signal case. The voxel-permutation experiment confirms that the detection
rate is reduced when we disrupt the spatial signal.  } \label{permute}
\end{figure}

To further assess the properties of \verb+GraphMM+, we  performed two permutation experiments leveraging
the ADNI-2 data.  In the first, we permuted the sample
labels of the 148 control subjects and 123 late MCI subjects, repeating for ten permuted sets.
On each permuted set, we applied various methods to detect differences.  All discoveries are false
discoveries in this null case. The left panel of Fig.~\ref{permute} shows that \verb+GraphMM+
and other methods are correctly recognizing the apparent  signals as being consistent with the null hypothesis.
The second permutation experiment retains the sample-grouping information, but permutes
the voxels within the brain slice on test.  This permutation disrupts both spatial
measurement dependencies and any spatial structure in the signal.  Since \verb+GraphMM+ is 
 leveraging spatially-coherent patterns in signal, we expect it to produce fewer statistically
significant findings in this voxel-permutation case.  The right panel of Fig.~\ref{permute}
shows this dampening of signal as we expect, when looking at the empirical cdf of computed
values $l_v=P(H_{0,v}|X,Y)$.

\subsection{ADNI-2 data analysis} \label{Application}

Using data from ADNI-2
we seek to evaluate the sensitivity of \verb+GraphMM+ in identifying significant
  differences between two disease stages (cognitively normal controls and late MCI), and also 
  assess the extent to which our findings are corroborated by 
known results on aging and Alzheimer's disease.
The methodology detects locations (i.e., voxels) where the distribution of MRI-based gray matter 
intensities is significantly different between control subjects and subjects with late MCI.
 To keep the computational burden manageable, we applied \verb+GraphMM+ 
to 2D image slices in the coronal direction, instead of processing the entire 3D image volume at once.
For comparison, we applied  Statistical non-parametric Mapping toolbox using  Matlab, \texttt{SnPM}, 
which is a popular image analysis method used in neuroscience, and $q$-value with adaptive shrinkage using R, 
\texttt{ashr}, which represents an advanced voxel-specific empirical-Bayes method.

\begin{figure}[p]
\center
\includegraphics[width=1\textwidth]{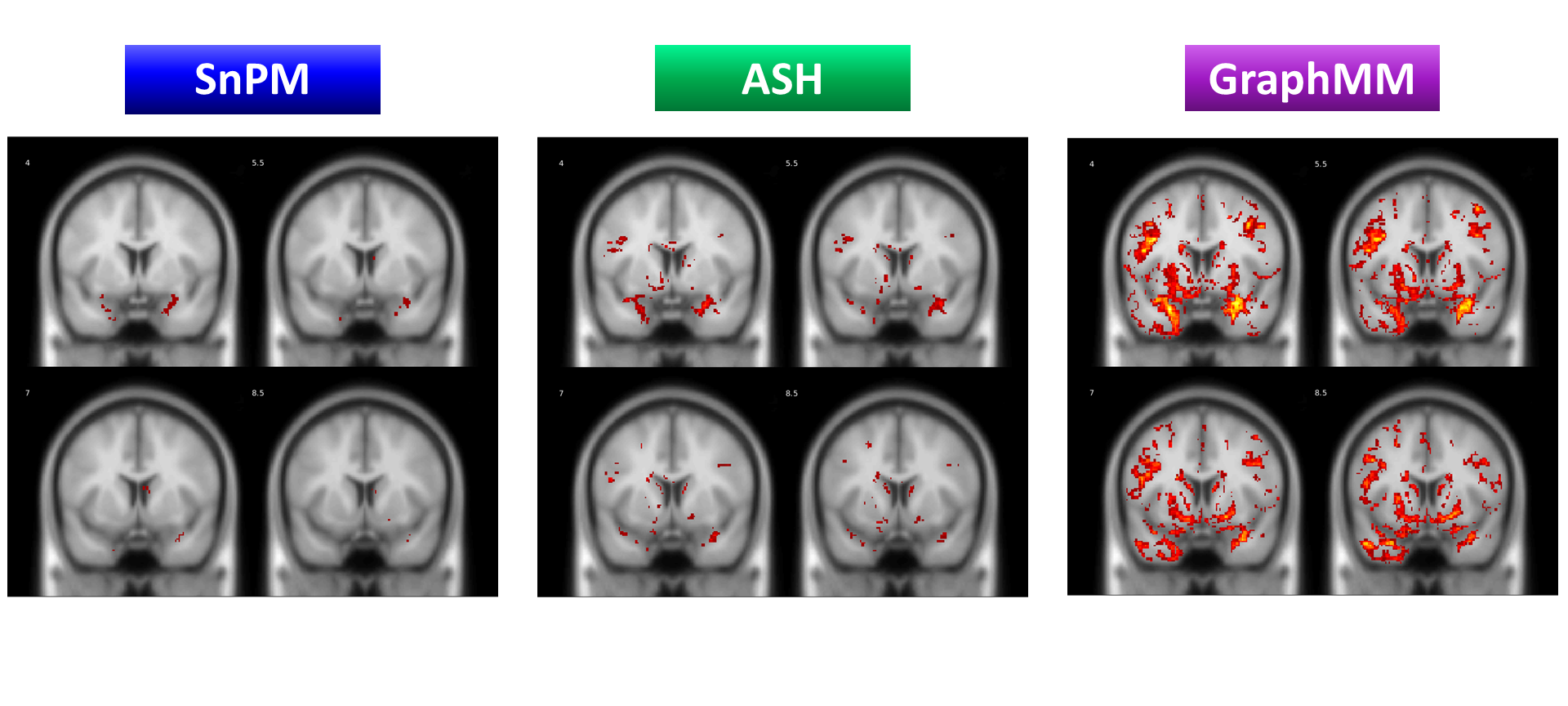}
\vspace{-0.4in}
\caption{Figure shows significantly different voxels at 5\% FDR (colored area) for 4 coronal slices, found by
Statistical non-parametric mapping (SnPM), adaptive shrinkage (ASH) and the proposed GraphMM.
} \label{4slices}
\end{figure}

\begin{figure}[p]
\center
\includegraphics[width=\textwidth]{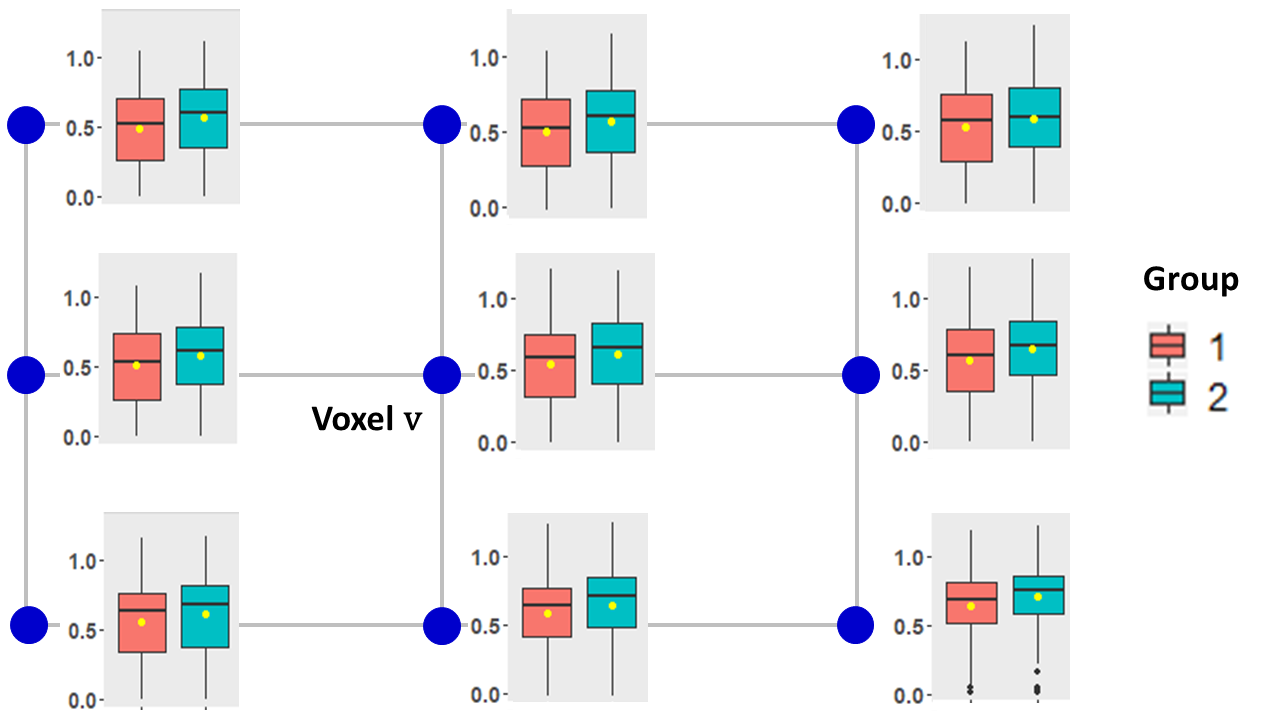}
\caption{Boxplots for voxel $v$ at coordinates $(x = 31, y = 53, z = 23)$ and its neighbors. Voxel $v$ is 
altered according to GraphMM but not according to SnPM or $q$-value.  Similar shifts nearby $v$ lead to the
increased evidence reported by GraphMM.
} \label{Boxplot}
\end{figure}

 Fig.~\ref{4slices} shows a representative example output 
 for a montage of 4 coronal slices extracted from the 3D image volume.
The color bar (red to yellow), for each method presented,
is a surrogate for the strength of some score describing the group-level difference:
for instance, for SnPM, the color is scaled based on adjusted $p$-values, for the $q$-value method, it is scaled
based on $q$-values, whereas for \verb+GraphMM+, the color is scaled based on local false-discovery rates $l_v$.
While the regions reported as significantly different between controls and late MCI have some overlap
between the different methods, \verb+GraphMM+ is able to identify many more significantly different voxels compared to baseline methods,
at various FDR thresholds (Supplementary Material, Fig. S4).
A closer inspection of one case is informative (Fig. \ref{Boxplot}).
Voxel $v$ at coordinates $(x = 31, y = 53, z = 23)$ is not found to be different between control and late MCI 
according to SnPM (adjusted $p$-value = 0.578) or the ASH $q$-value method ($q$-value = 0.138). But when we look at the results
provided by \verb+GraphMM+, the local FDR is $0.001$.  
The increased sensitivity of \verb+GraphMM+ may come from its  leveraging the consistent pattern of shifts 
among neighboring voxels.

A statistical measure often reported in the neuroimaging literature is the size of
spatially connected sets of significantly altered voxels in the 3D lattice (so-called significant clusters). 
The rationale is that stray (salt and pepper) voxels reported as significantly different may be  more likely to be an artifact compared
to a group of anatomically clustered voxels.
 We find that \verb+GraphMM+ performs favorably relative to the baseline methods in that it consistently reports
larger significant clusters (Supplementary Material, Fig. S5).

To provide some neuroscientific  interpretation of the statistical findings, 
we use the Matlab package \emph{xjview} to link
anatomical information associated with significantly altered voxels~\citep{aal}.
Results for the top 15 brain regions  are summarized in Table~\ref{brainregion}. 
We see that \verb+GraphMM+ discovers all the brain regions found by SnPM, with many more significant voxels in
each region. The only exception is the hippocampus, 
where both methods identify a large number of voxels but \verb+GraphMM+ finds fewer significant voxels than SnPM.
In addition, there are regions revealed to be significant by \verb+GraphMM+ but not by SnPM, including the precentral gyrus,
middle frontal gyrus, inferior frontal gyrus opercular, insular, anterior cingulate, and supramarginal gyrus, which are
relevant in the aging and AD literature. 
\verb+GraphMM+ consolidates known alterations between cognitively normal  and late-stage MCI and reveals 
potentially important new findings.

\section{Discussion}

Mass univariate testing is the dominant approach to detect statistically significant changes in comparative
brain-imaging studies (e.g., \cite{Groppe2011}).
In such, a classical testing procedure, like the test of a contrast in
a regression model,  
is applied in parallel over all testing units (voxels), leading to a large number of univariate test statistics 
and  p-values.   Subsequently, significant voxels are identified through some filter, such as the Benjamini-Hochberg (BH)
procedure, that aims to control the false-discovery rate.
The approach can be very effective and has supported numerous important
applied studies of brain function.
 In structural magnetic resonance image  studies of Alzheimer's disease progression, such mass
univariate testing has failed in some cases to reveal subtle structural changes  between phenotypically distinct
patient populations.   The underlying problem is limited statistical power
for relatively small effects, even with possibly hundreds of subjects per group.   
Power may be recovered by 
empirical Bayes procedures that leverage various properties of the collection of tests.  The proposed \verb+GraphMM+ method
recognizes simplified parameter states when they exist among graphically related testing units. 
We deploy \verb+GraphMM+ locally in the system-defining graph 
by separately processing a small subgraph for each testing unit, while allowing
hyper-parameters to be estimated globally from all testing units.  Essentially we provide an explicit and flexible
joint probability model for all data on each subgraph (Equation \ref{lkh}).   
The model entails a discrete parameter state on this 
subgraph, which describes how the nodes on the subgraph are partitioned into blocks, and whether or not each block 
is shifted between the two sampling conditions being compared.  By deriving local FDR computations on a relatively
small subgraph for each testing unit, we simplify computations and we share  perhaps the most
relevant information that is external to that testing unit.
Numerical experiments confirm the control of FDR and the beneficial power properties of \verb+GraphMM+, whether the
model specificaton is valid or violated in various ways.  The methodology also reveals potentially interesting
brain regions that exhibit significantly different structure between normal subjects and those suffering mild cognitive
impairment.

The Dirichlet process mixture (DPM) model also 
entails a clustering of the inference units, with units in the same cluster block if (and only if) they share the same parameter values.   The DPM model has been effective at representing heterogeneity in a system of parameters (e.g., \cite{Muller2004}), and in improving sensitivity in large-scale testing (e.g., \cite{Dahl_Newton_2007}, \cite{Dahl2008}). Benefits typically come at a high computational cost, since in principle the posterior summaries require averaging over all partitions of the units  (e.g., \cite{Blei2006}). There are also modeling costs: DPM's usually have a product-partition form in which the likelihood function factors as a product over blocks of the partition (\cite{Hartigan1990}).  We observe that independence between blocks 
is violated in brain-image data in a way that may  lead to inflation of the actual false-discovery rate over a target value.

Graphs have been widely used for modeling data arising in various fields of science. 
In the present work, vertices of the graph correspond to variables in a data set  and
the undirected edges convey relational information about the connected variables, due to
associations with the context of the data set, such as  temporal, functional, spatial, or anatomical
information.  The graphs we consider constitute 
an auxiliary part of observed data.  For clarity,  these graphs may or may
not have anything to do with undirected graphical representations of the dependence in a joint distribution (e.g. \cite{GraphicalModel}), as in the graphical models literature.  For us, the graph serves to constrain patterns in the expected values of measurements.  By limiting changes in expected values over the graph, we aim to capture
low complexity of the system.  An alternative way to model low-complexity 
is through  smoothed, bandlimited signals (e.g., \cite{Antonio2018}, \cite{Chen2016}).  Comparisons between the
approaches are warranted.  We have advanced the idea of 
latent graph-respecting partitions that constrain expected values into low-dimensional
space.  Figure~S3 in the Supplementary Material investigates benefits of the graph-respecting assumption on 
posterior concentration, and supports the treatment of this constraint as having a regularizing effect.
The partition is paired with a vector of block-specific change indicators to convey the discrete part of
the modeling specification.  We used a uniform distribution over graph-respecting partitions in our numerical
experiments, and have also considered more generally the distribution found by conditioning a product partition model (PPM) to be
graph-respecting.   In either case, 
two vertices that are nearby on the graph are more likely to share expected values, in contrast to the 
exchangeability inherent in most partition models.   Graph restriction greatly
reduces the space of partitions; we enumerated all such partitions in our proposed graph-local computations.
 When the generative situation is similarly graph restricted, we expect
improved statistical properties; but we also showed that false-discovery rates are controlled even if the generative
situation is not graph respecting.   Special cases of graph-restricted partitions have been studied by others,
including \cite{Page2016} for lattice graphs, \cite{Caron2009} for decomposable graphs, and \cite{Blei2011} for graphs
based on distance metrics.  When $G$ is a complete graph there is no restriction and all partitions have positive mass. 
When $G$ is a line graph the graph-respecting partition  model matches \cite{Barry1992} for change-point analysis.

\section*{Acknowledgements} This research was supported in part by NIH grant 1U54AI117924 to the University 
of Wisconsin Center for Predictive Computational Phenotyping.  The authors were also supported in part 
by NIH R01 AG040396, NSF CAREER award RI 1252725, and NSF award 1740707 to the University of Wisconsin Institute
for the Foundations of Data Science. We are grateful to Prof. Barbara Bendlin (Wisconsin ADRC, University of Wisconsin) for various
discussions and in particular, for her help in evaluating the results on brain imaging data. 

\small
\bibliographystyle{biorefs}
\bibliography{References}
\normalsize


\section*{Figures and Tables}

\begin{table}[p]
{ \caption{Brain regions with significant (5\% FDR) change in gray matter volume found by GraphMM.}  \label{brainregion}
\footnotesize
 \begin{tabular}{clcl}
    \toprule
       No.    & \begin{minipage}{0.17\textwidth} Brain region \end{minipage} & \begin{minipage}{0.2\textwidth}\vspace{0.01in} { ~~~~~~~\# Voxels} \vspace{-0.09in} \\  \rule{\textwidth}{.6pt} \vspace{0.06in}
       GraphMM ~~~ SnPM
       \end{minipage} & ~~~~~~~~~~~~~~~~~~~~ Neurological function  \\ \midrule
    1 & Hippocampus & \begin{minipage}{0.15\textwidth}  \begin{tabular}{ll}
       1411 & 1646
       \end{tabular}    \end{minipage} &\begin{minipage}{0.55\textwidth}     Receives and consolidates new memory about experienced events, allowing for establishment of long-term memories. [\cite{hippocampus}]  \end{minipage}  \\ \midrule
    2 & \begin{minipage}{0.19\textwidth} {Parahippocampal}\\ {Gyrus}    \end{minipage} & \begin{minipage}{0.15\textwidth}  \begin{tabular}{ll}
       1008 & 410
       \end{tabular}    \end{minipage}& \begin{minipage}{0.55\textwidth}
     Involved in episodic memory and visuospatial processing [\cite{parahippocampal}]. \end{minipage} \\ \midrule
    3 & Amygdala & \begin{minipage}{0.15\textwidth}  \begin{tabular}{ll}
       710 & 365
       \end{tabular}    \end{minipage}& \begin{minipage}{0.55\textwidth}
                        Plays an essential role in the processing and memorizing of emotional reactions [\cite{amygdala}] \end{minipage} \\ \midrule
        4 & \begin{minipage}{0.19\textwidth} {Temporal Gyrus}\\{ (superior, middle and inferior)}       \end{minipage}& \begin{minipage}{0.15\textwidth}  \begin{tabular}{ll}
       2031 & 287
       \end{tabular}    \end{minipage}&\begin{minipage}{0.55\textwidth}
   Involved in various cognitive processes, including language and semantic memory processing (middle) as well as visual perception (inferior) and sound processing (superior) [\cite{temporal_gyrus_mid_inf}, \cite{temporal_gyrus_sup}]       \end{minipage}\\ \midrule
                        5 & { Putamen} &\begin{minipage}{0.15\textwidth}  \begin{tabular}{ll}
       793 & 12
       \end{tabular}    \end{minipage} &\begin{minipage}{0.55\textwidth}
                        {Linked to  various types of motor behaviors, including motor planning, learning, and execution.[\cite{putamen}].}
                        \end{minipage} \\ \midrule
                        6 & {Fusiform Gyrus} & \begin{minipage}{0.15\textwidth}  \begin{tabular}{ll}
       735 & 308
       \end{tabular}    \end{minipage} & \begin{minipage}{0.55\textwidth}
                        {Influence various neurological phenomena including face perception, object recognition, and reading [\cite{fusiform}].}
                        \end{minipage} \\ \midrule
                        7 & \begin{minipage}{0.19\textwidth}
                        {Temporal Pole}\\{ (superior, middle)}
                        \end{minipage}   & \begin{minipage}{0.15\textwidth}  \begin{tabular}{ll}
       882 & 74
       \end{tabular}    \end{minipage} & \begin{minipage}{0.55\textwidth}
                        {Involved with multimodal analysis, especially in social and emotional processing. [\cite{temporal_pole}].}
                        \end{minipage}\\ \midrule
                        8 & {Precentral Gyrus}  &\begin{minipage}{0.15\textwidth}  \begin{tabular}{ll}
       829 & 0
       \end{tabular}    \end{minipage}& \begin{minipage}{0.55\textwidth}
                        {Consists of primary motor area, controlling body's movements. [\cite{precentral}].}
                        \end{minipage}\\ \midrule
                        9 & \begin{minipage}{0.19\textwidth}
                        {Middle Frontal}\\ Gyrus
                        \end{minipage}   & \begin{minipage}{0.15\textwidth}  \begin{tabular}{ll}
       635 & 0
       \end{tabular}    \end{minipage}& \begin{minipage}{0.55\textwidth}
                        {Plays essential role in attentional reorienting. [\cite{mid_frontal}].}
                        \end{minipage}\\ \midrule
                        10 & \begin{minipage}{0.19\textwidth}
                        {Inferior Frontal}\\{Gyrus Opercular}\end{minipage}   & \begin{minipage}{0.15\textwidth}  \begin{tabular}{ll}
       573 & 0
       \end{tabular}    \end{minipage}& \begin{minipage}{0.55\textwidth}
                        {Linked to language processing and speech production. [\cite{inf_frontal}].}
                        \end{minipage}\\ \midrule
                        11 & {Calcarine} & \begin{minipage}{0.15\textwidth}  \begin{tabular}{ll}
       381 & 22
       \end{tabular}    \end{minipage} &\begin{minipage}{0.55\textwidth}
                        {Where the primary visual cortex is concentrated, processes visual information. [\cite{calcarine}].}
                        \end{minipage}\\ \midrule
                        12 & {Caudate} & \begin{minipage}{0.15\textwidth}  \begin{tabular}{ll}
       437 & 46
       \end{tabular}    \end{minipage} & \begin{minipage}{0.55\textwidth}
                        {Plays essential roles in motor processes and a variety of executive, goal-directed behaviours [\cite{caudate}].}                       \end{minipage} \\ \midrule
                        13 & {Insular}  & \begin{minipage}{0.15\textwidth}  \begin{tabular}{ll}
       275 & 0
       \end{tabular}    \end{minipage} & \begin{minipage}{0.55\textwidth}
                        {Involved in consciousness, emotion and the regulation of the body's homeostasis [\cite{insular}].}
                        \end{minipage}\\ \midrule
                        14 & \begin{minipage}{0.19\textwidth} {Anterior \\ Cingulate}\end{minipage} & \begin{minipage}{0.15\textwidth}  \begin{tabular}{ll}
       260 & 0
       \end{tabular}    \end{minipage} & \begin{minipage}{0.55\textwidth}
                        {Plays a major role in mediating cognitive influences on emotion. [\cite{anterior}].}
                        \end{minipage}\\ \midrule
                        15 & \begin{minipage}{0.19\textwidth} {Supramarginal \\ Gyrus} \end{minipage} &\begin{minipage}{0.15\textwidth}  \begin{tabular}{ll}
       225 & 0
       \end{tabular}    \end{minipage} & \begin{minipage}{0.55\textwidth}
                        {Linked to phonological processing and emotional responses. [\cite{supra}].}                    \end{minipage}  \\
  \bottomrule
  \end{tabular}
\vspace{0.5in}
}

\end{table}

\end{document}